\documentclass[preprint,aps]{revtex4}
\usepackage[latin9]{luainputenc}
\usepackage{float}
\usepackage{amsmath}
\usepackage{color}
\usepackage{amssymb}
\usepackage{graphicx}
\usepackage{setspace}

\makeatletter

\@ifundefined{textcolor}{}
{%
 \definecolor{BLACK}{gray}{0}
 \definecolor{WHITE}{gray}{1}
 \definecolor{RED}{rgb}{1,0,0}
 \definecolor{GREEN}{rgb}{0,1,0}
 \definecolor{BLUE}{rgb}{0,0,1}
 \definecolor{CYAN}{cmyk}{1,0,0,0}
 \definecolor{MAGENTA}{cmyk}{0,1,0,0}
 \definecolor{YELLOW}{cmyk}{0,0,1,0}
}

\usepackage{amsfonts}
\usepackage{bm}
\usepackage[dvips]{epsfig}
\usepackage[dvips]{graphicx}
\usepackage{float}
\usepackage{dcolumn}
\usepackage{ifpdf}
\usepackage{multirow}

\newcommand{\sech}{\mathrm{sech}}
\newcommand{\be}{\begin{equation}}
\newcommand{\ee}{\end{equation}}
\newcommand{\bes}{\begin{subequations}}
\newcommand{\ees}{\end{subequations}}
\newcommand{\ben}{\begin{eqnarray}}
\newcommand{\een}{\end{eqnarray}}

\makeatother

\begin{document}

\title{Kink scattering in the presence of geometric constrictions}
\author{Jo\~ao G. F. Campos$^{1}$, Fabiano C. Simas$^{2,3}$, D. Bazeia$^{4}$}
\email{joaogfc@gmail.com, fc.simas@ufma.br, bazeia@fisica.ufpb.br}

%\selectlanguage{english}

\affiliation{
$^1$ Departamento de F\'isica, Universidade Federal de Pernambuco\\ Av. Prof. Moraes Rego, 1235, 50670-901, Recife, PE, Brazil\\
$^2$ Programa de P\'os-Gradua\c c\~ao em F\'\i sica, Universidade Federal do Maranh\~ao\\Campus Universit\'ario do Bacanga, 65085-580, S\~ao Lu\'\i s, Maranh\~ao, Brazil
\\$^3$ Centro de Ci\^encias de Chapadinha-CCCh, Universidade Federal do Maranh\~ao\\65500-000, Chapadinha, Maranh\~ao, Brazil\\
$^4$ Departamento de F\'isica, Universidade Federal da Para\'iba\\ 58051-970, Jo\~ao Pessoa, PB, Brazil
}

\begin{abstract}

We investigate kink-antikink collisions in a model characterized by two scalar fields in the presence of geometric constrictions. The model includes an auxiliary function that modifies the kinematics associated with one of the two fields. An important fact is that one of the fields can be solved independently, being responsible for changing the internal structure of the second one. We performed several collisions and observed the presence of resonance windows for small values of the parameters. Furthermore, we have been able to show the alternation between the appearance of oscillating pulses, as well as the annihilation and formation of kink-antikink pairs when the geometric constriction is more pronounced. The study of kink dynamics in models with geometric constrictions is connected with issues of interest such as domain wall formation and magnetization at the manometric scale.

\end{abstract}

\maketitle

%%%%%%%%%%%%%%%%%%%%%%%%%%%%%%%%%%%%%%%%%%%%%%%%%%%%%%%

\section{ Introduction }

%%%%%%%%%%%%%%%%%%%%%%%%%%%%%%%%%%%%%%%%%%%%%%%%%%%%%%%

Topological defects have been the subject of extensive research over the years due to their complex physical behavior and potential applications. In particular, defects appear in high energy physics \cite{masu,vacha} and also in the brane scenario, allowing the internal structure of the brane to be controlled \cite{balo}. Additionally, defects can also be found in condensed matter physics \cite{poga,chdo}. In the interesting paper \cite{jab}, for instance, the authors investigated the configuration of a wall experimentally and by micromagnetic simulations in the presence of constrained geometries. There, they identified an interesting behavior, the division of the wall into a two-kink structure.

Kinks are topological defects generated by real scalar fields in $(1,1)$ spacetime dimensions. The presence of two or more scalar fields brings higher complexity and may lead to interesting developments. Over the last decades, collisions between kinks in theories with scalar fields have attracted great interest. Most especially, in the set of works  \cite{anninos,sugi,camp1}, the authors demonstrated the appearance of resonance windows and a chaotic structure created by kink-antikink collisions. In the present days, we know that the mechanism of resonant energy exchange between translational and vibrational modes is responsible for the appearance of this structure, with a standard linear perturbation theory used to obtain the internal modes. Additionally, there are many other papers that discuss kink collision in various models as well; see, e.g., Refs. \cite{alfaba,begazl,chdedegakesa,gamoja,momo} and references therein. We can also mention works such as the sine-Gordon model \cite{blms,masd,pmrg,bgmsa}, non-integrable $\phi^4$ \cite{cape,dohamerosh,asmoraebsa} and $\phi^6$ \cite{dorosh,weig,mogasadmja} models, and the scattering of wobbling kinks \cite{alquni,jomo,jomo1}.

As we increase the number of scalar fields, the field theory acquires much higher complexity. Interestingly, field theories of two coupled scalar fields can be used to describe Bloch walls \cite{bazeia2022} and Bloch branes \cite{bazeia2004}. The scattering between topological and non-topological structures in such theories is a very rich field of research with highly nontrivial results. In \cite{alonso}, the authors considered such scattering in a model containing one degenerate vacuum. Cases containing two \cite{alonso2,alonso3}, three \cite{algomato} and four \cite{Halavanau,simas2022} have also been considered. Moreover, the scenario where one of the fields is in its quantum vacuum was investigated in Refs.~\cite{maevvaza,mavacha}.

A subject that has been widely studied in recent years is the appearance of spectral walls \cite{adrowe,camoquweza}, which correspond to a formed obstacle in kink-antikink collisions, that arises from the transition from normal mode to continuous mode. In particular, it was established in Ref. \cite{adrowe1} that the modes connected to antikink-kink collisions are responsible for the appearance of the thick spectral walls. Furthermore, in models with two scalar fields, the spectral wall phenomenon in kink dynamics can be observed \cite{adolroqwza}.

A distinct class of multifield models was investigated recently in Ref. \cite{blm}, with a special focus on the case with two real scalar fields. There, the kinetic component of the model's Lagrangian was modified, resulting in the presence of an interesting internal structure. In Ref. \cite{jab}, for instance, the main ingredient for the appearance of the two-kink solution is the presence of the geometric constriction, which was shown to be directly connected with the kinetic modification introduced in \cite{blm}. Another work in which the use of constrained geometries is important can be found in Ref. \cite{hiatma}, where the authors discuss Ni$_{81}$Fe$_{19}$ thin film magnetic structures that exhibit domain wall nucleation in clearly defined nanoscale constrictions. Also, in Ref. \cite{clahu} the process of fabricating nanostructures for constrained domain walls was reported. Moreover, it is worth noting the investigation of confined domain walls in magnetic nanotubes \cite{chego}. There, the simulations revealed that the magnetization structure of the constrained domain walls is directly proportional to the size of the tubes. Furthermore, one of the magnetization components was shown to exhibit two-kink behavior.

The study of kinks in systems with two scalar fields has been a topic of great interest in theoretical physics for several decades. The dynamics of such kinks can become especially fascinating under the presence of geometric constriction, and they might be extremely important in the formation of domain walls. The results of studying the collision of kinks in a model with kinetic modifications deserve further attention, since the presence of geometric constrictions produces kink profiles with internal structure, consequently changing the physical characteristics of the scattering. For instance, research on charge transfer at the nanoscale results in a current with two-kink properties \cite{thivan}. Additionally, it was investigated in Ref. \cite{fermion} that the presence of geometric modifications affects the behavior of fermions.

The main goal of the research is to comprehend how the dynamics of kinks in a model with two scalar fields can be affected by the presence of geometric constrictions. In order to implement the investigation, we organize the work as follows. In the next section, we present the model and investigate its kink-like configurations, where a function is introduced that modifies the kinematics of one of the scalar fields. In Sec.~\ref{sec:stab}, we analyze the behavior of perturbations around the static solutions. In Sec. \ref{sec3}, we present an extensive numerical analysis of the kink-antikink scattering for several values of the pertinent parameters that control specific properties of the system. We conclude the study in Sec. \ref{sec4}, where we add comments and suggestions for future work.

%%%%%%%%%%%%%%%%%%%%%%%%%%%%%%%%%%%%%%%%%%%%%%%%%%%%%%%%%%%%%%%%%%%%%%%%%%%

\section { The Model } \label{sec2}

%%%%%%%%%%%%%%%%%%%%%%%%%%%%%%%%%%%%%%%%%%%%%%%%%%%%%%%%%%%%%%%%%%%%%%%%%%%

We present a field theoretical model in $(1,1)$ dimensions with two scalar fields, $\phi$ and $\chi$. The Lagrangian density is described by
\be
{\cal{L}} = \frac12 f(\chi) \partial_{\mu} \phi \partial^{\mu} \phi +\frac12 \partial_{\mu} \chi \partial^{\mu} \chi   - V(\phi, \chi).
\label{lagrangian}
\ee
This type of construction has already been investigated in Ref \cite{blm,bbm}. The function $f(\chi)$ only depends on $\chi$, and it is in principle an arbitrary non-negative function that is in charge of changing the kinematic component of the field $\phi$. This function denotes geometrical constrictions that change the internal structure of the kink-like structures. The real scalar fields are also coupled through the potential $V(\phi,\chi)$. We follow Refs. \cite{blm,bbm}, and consider the potential in the form
\be
V(\phi,\chi)=\frac12 \frac{W_\phi^2}{f(\chi)} + \frac12 W_\chi^2,
\ee
where $W=W(\phi,\chi)$, and $W_{\phi}=\partial W/\partial \phi$ and $W_{\chi}=\partial W/\partial \chi$. 

\begin{figure}
	\includegraphics[{angle=0,width=8cm,height=6cm}]{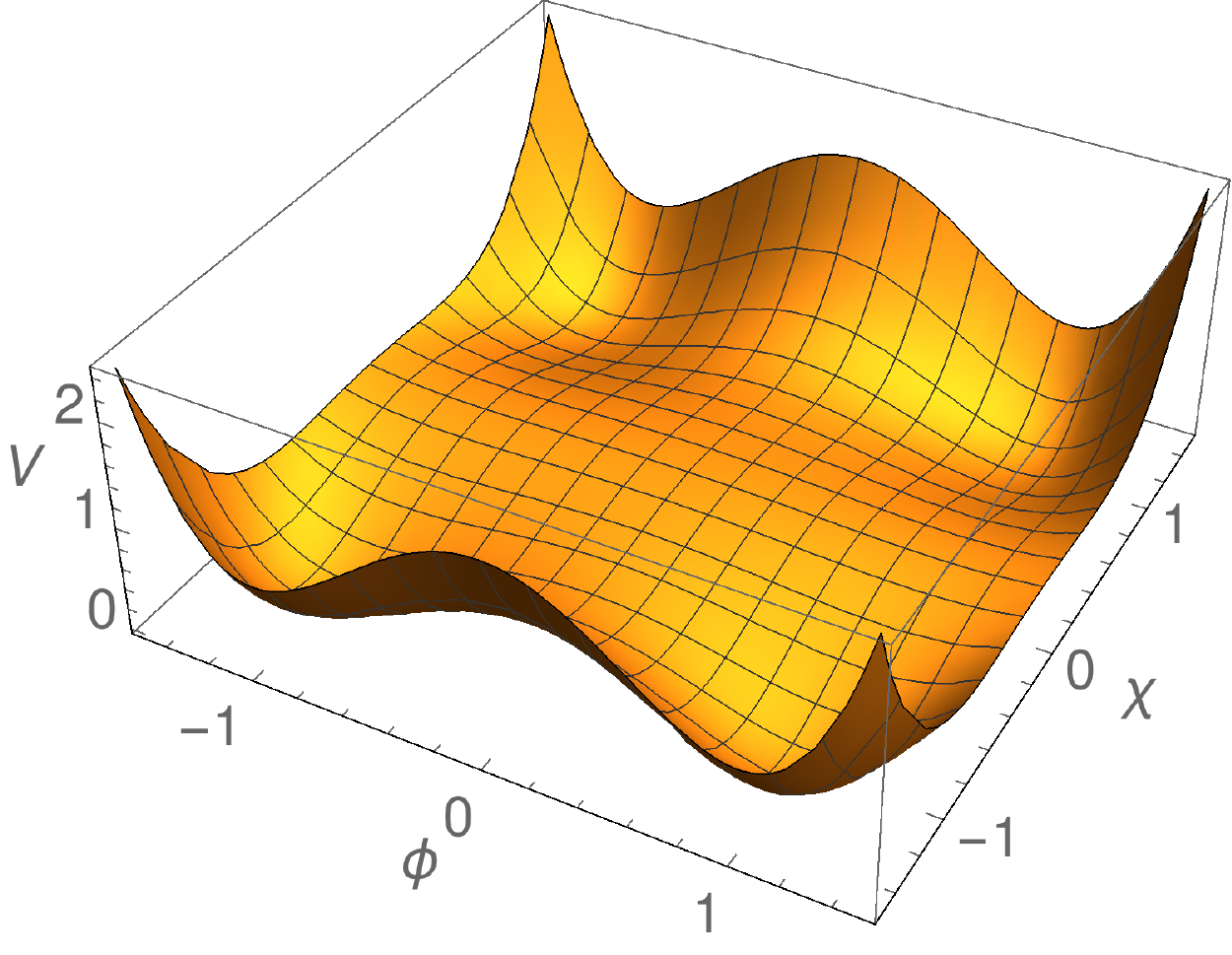}
        \includegraphics[{angle=0,width=8cm,height=6cm}]{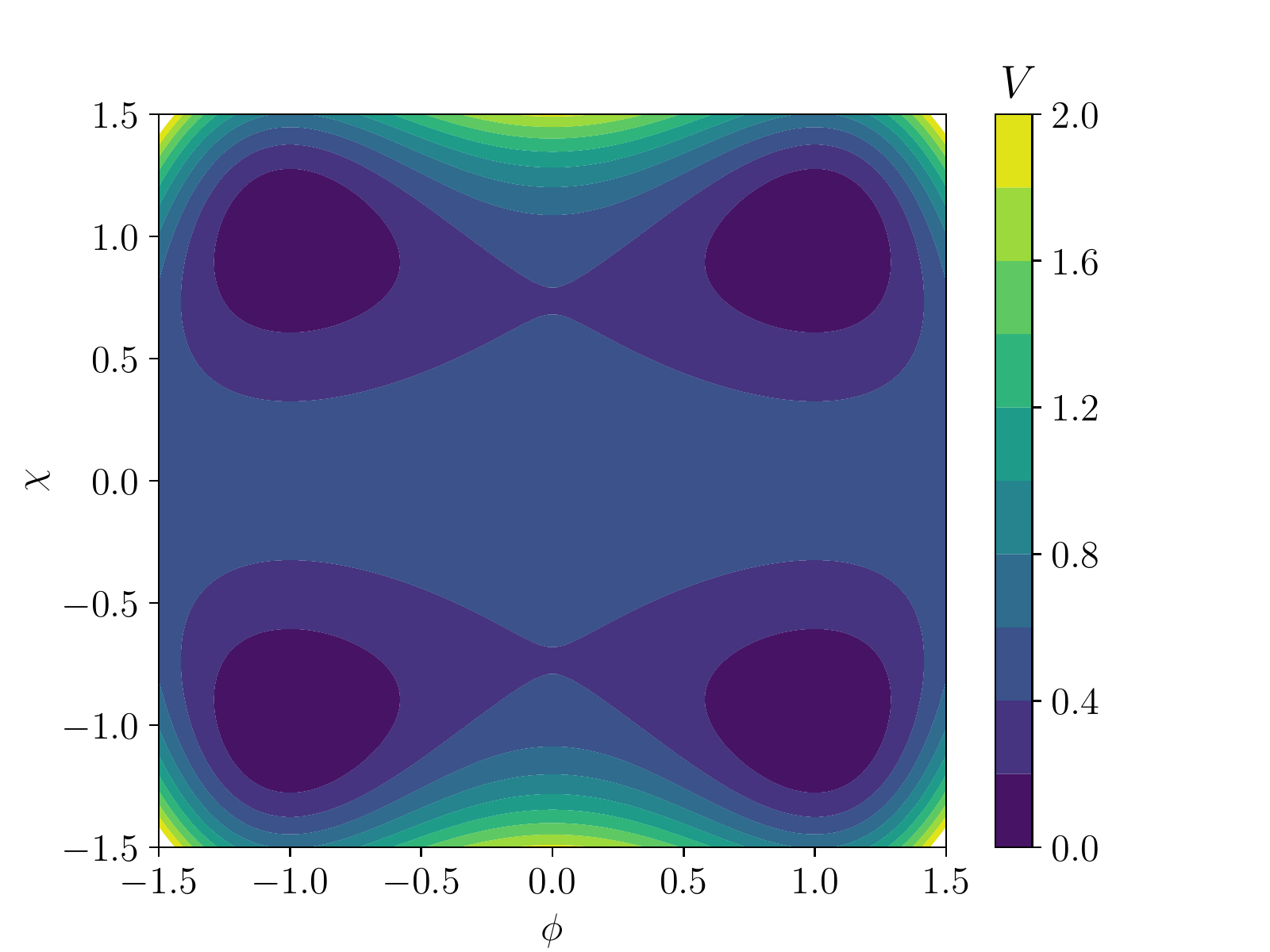}
        \caption{Potential surface as a surface and contour plots.}
	\label{pot-surf}
\end{figure}

The equations of motion for the scalar fields are given by
\begin{eqnarray}
\frac{\partial f}{\partial t}\frac{\partial \phi}{\partial t} -\frac{\partial f}{\partial x}\frac{\partial \phi}{\partial x} + f\frac{\partial^2 \phi}{\partial t^2} - f\frac{\partial^2 \phi}{\partial x^2}+\frac{\partial V(\phi,\chi)}{\partial\phi} & = & 0,
\label{eqm1}\\
\frac{\partial^2 \chi}{\partial t^2}-\frac{\partial^2 \chi}{\partial x^2} + \frac12 \frac{d f}{d \chi}\bigg[ \bigg( \frac{\partial \phi}{\partial x} \bigg)^2 - \bigg( \frac{\partial \phi}{\partial t} \bigg)^2 \bigg] + \frac{\partial V(\phi,\chi)}{\partial\chi} & = & 0.
\label{eqm2}
\end{eqnarray}
The energy density can be expressed using the arbitrary function W; it is given by
\be
\mathcal{\rho} = \frac12 f(\chi) \biggl( \frac{d\phi}{dx} \mp \frac{W_\phi}{f(\chi)} \biggr)^2 + \frac 12 \biggl( \frac{d\chi}{dx} \mp W_\chi \biggr)^2 \pm \frac{dW}{dx}.
\ee
Consequently, we are able to write the first-order, or Bogomol'nyi-Prasad-Sommerfield (BPS), equations
\be
 \frac{d\phi}{dx} = \pm \frac{W_{\phi}(\phi,\chi)}{f(\chi)}, \,\,\,  \frac{d\chi}{dx} = \pm W_\chi(\phi,\chi).
\label{firtorder}
\ee
These equations minimize the energy to $E = |W( \phi(+\infty), \chi(+\infty)) -  W( \phi(-\infty), \chi(-\infty))|$. Therefore, the $f(\chi)$ function does not contribute to the energy, which depends only on the asymptotic values of the field configurations. In this work, let us use the potential given by Ref. \cite{blm}
\be
V(\phi, \chi) = \frac{1}{2f(\chi)} (1-\phi^2)^2 + \frac12 \alpha^2(1-\chi^2)^2,
\label{pot}
\ee
where $\alpha$ is a real and non-negative parameter and for $\chi=\pm 1$ and $f=1$, it is possible to recover the standard $\phi^4$ model. The potential surface is shown in Fig.~\ref{pot-surf} and engenders kinks and antikinks, which are solutions that interpolate two distinct minima. As there exist four minima, there are in total 6 topological sectors, engendering 12 families of topological solutions. They are related by the $Z_2\times Z_2$ symmetry of the model. There are four symmetrically related solutions which are $\phi^4$ kinks and trivial in $\chi$.  Similarly, there are also four symmetrically related solutions which are $\chi^4$ kinks with trivial $\phi$. Finally, there are four families of solutions that are nontrivial in both fields, which are also symmetrically related. Notably, the last ones result in analytical solutions with internal structure, as we will show below.

%%%%%%%%%%%%%%%%%%%%%%%%%%%%%%%%%%%%%%%%%%%%%%%%%%%%%%%%%%%%%%%%%%
%%%%%%%%%%%%%%%%%%%%%%%%%%%%%%%%%%%%%%%%%%%%%%%%%%%%%%%%%%%%%%%%%%

Now, we can rewrite the first-order equations given by
\begin{eqnarray}\label{eqorder1}
\frac{d\phi}{dx} &=& \pm \frac{1}{f(\chi)}(1 - \phi^2), \\
\frac{d\chi}{dx} &=& \pm \alpha(1-\chi^2)
\label{eqorder2},
\end{eqnarray}
the plus and minus signs corresponding to kinks and antikinks. The Eq. \eqref{eqorder2} for $\chi$ 
is independent of $\phi$ and can be readily solved. It leads to a kink-like solution connecting the minima $\chi_-=-1$ to $\chi_+=+1$
\begin{eqnarray}\label{chi}
    \chi_K(x) = \tanh(\alpha x).
\end{eqnarray}
Due to the translational symmetry, we may set the integration constant to zero. Similarly, the minus sign in the antikink-like solution connects the minimum $\chi_+=+1$ to $\chi_-=-1$.
%%%%%%%%%%%%%%%%%%%%%%%%%%%%%%%%%%%%%%%%%%%%%%%%%%%%%%%%%%%%%%%%%%
\begin{figure}
	\includegraphics[{angle=0,width=16cm,height=6cm}]{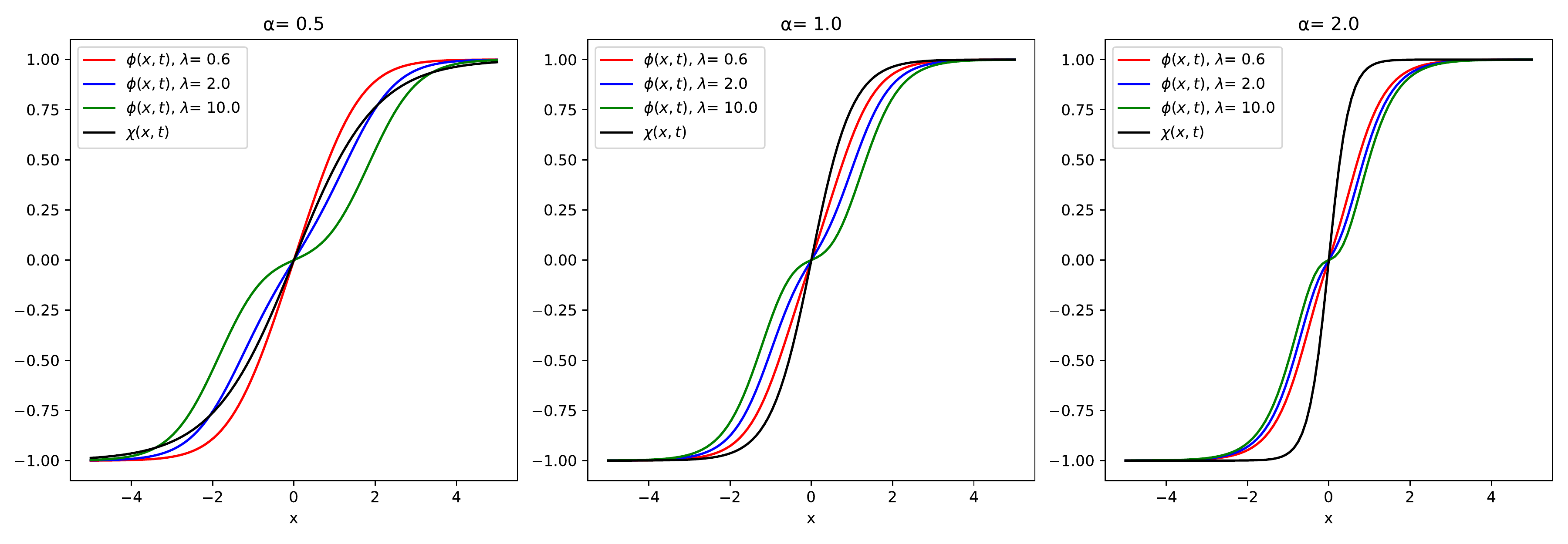}
	\caption{Kink solution for $\chi(x)$ (black solid) and for $\phi(x)$ with $\lambda=0.6$ (red), $\lambda=2$ (blue) and $\lambda=10$ (green) for (a) $\alpha=0.5$, (b) $\alpha=1.0$ and (c) $\alpha=2.0$. We fix $y_0=0$.}
	\label{sol1}
\end{figure}
%%%%%%%%%%%%%%%%%%%%%%%%%%%%%%%%%%%%%%%%%%%%%%%%%%%%%%%%%%%%%%%%%%

In this work, we introduce a function that affects the internal structure of the $\phi$ field
\begin{eqnarray}\label{F}
    f(\chi) = \frac{1+\lambda}{1+\lambda\chi^2},
\end{eqnarray}
where $\lambda$ is a non-negative real parameter. The function written in this manner changes the center of the $\phi$ field smoothly, as we vary $\lambda$; however, it does not change the tail of the solution \cite{bbm}. 
Next, we perform a change of variable $x\to y(x)$ in the first-order equation \eqref{eqorder1} to transform it into $d\phi/dy=\pm(1-\phi^2)$, which is solved by the kink-like configuration $\phi(y)=\tanh(y)$. We then use \eqref{F} to get to the explicit $x-$dependent solution 
\begin{eqnarray}\label{phi}
    \phi_K(x) = \tanh\bigg( y_0 + x - \frac{\lambda}{\alpha(1+\lambda)}\tanh( \alpha x)  \bigg),
\end{eqnarray}
where $y_0$ is an integration constant, being responsible for shifting the center of the kink in the vertical axis \cite{fermion,bbm}. It is important to note that the standard solution can be recovered when $\lambda \to 0$. The kink profile for $\chi(x)$ and $\phi(x)$ is depicted in Fig. \ref{sol1} for some values of $\alpha$ and $\lambda$, and $y_0=0$. The solutions become more localized at the origin as $\alpha$ increases; see Ref. \cite{blm}. An interesting feature of the $\chi$ field is its independent character, and furthermore, it can capture the $\phi$ field, creating an internal structure. We notice that as the value of $\lambda$ increases, it contributes to the appearance of a null derivative at the center of the kink, leading to the formation of a two-kink structure. As seen in Ref. \cite{jab}, the field $\chi$ simulates the presence of geometric constrictions in the field $\phi$.

\begin{figure}
	\includegraphics[{angle=0,width=16cm,height=6cm}]{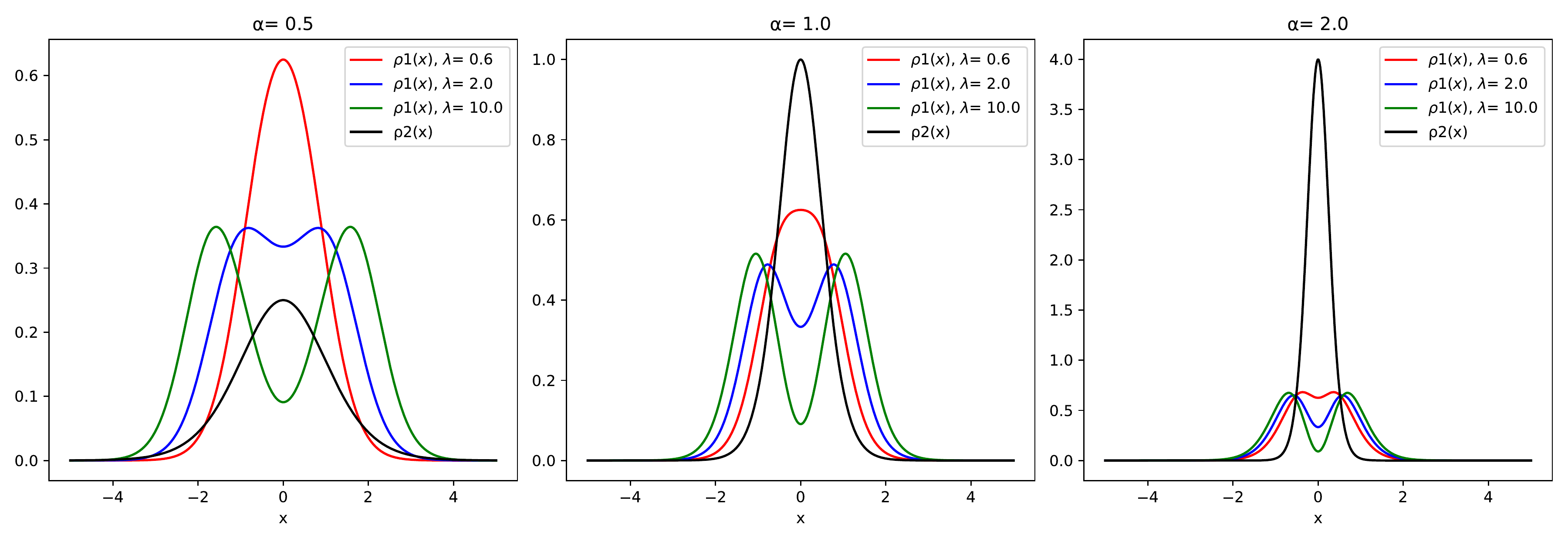}
	\caption{Energy density for $\rho_2$ (black solid) and for $\rho_1$ with $\lambda=0.6$ (red), $\lambda=2$ (blue) and $\lambda=10$ (green) for (a) $\alpha=0.5$, (b) $\alpha=1$ and (c) $\alpha=2$. We fix $y_0=0$.}
	\label{dens1}
\end{figure}

The energy density can be rewritten as the sum of two contributions, $\rho=\rho_1+\rho_2$, given by
\begin{eqnarray}
\rho_1 = \frac{W^2_{\phi}}{f(\chi)}, \quad \rho_2 =W^2_{\chi}.
\end{eqnarray}
In Fig. \ref{dens1}, we depict the energy density for some values of $\alpha$ and $\lambda$, and $y_0=0$. The structure becomes more localized as the parameter $\alpha$ is increased. Additionally, we observe that for small values of $\alpha$, the contribution of the $\phi$ field outweighs that of the $\chi$ field and shows a central maximum for small values of $\lambda$. We also notice that the formation of a plateau in the center of the solution (Fig. \ref{sol1}) contributes to the appearance of two points of maximum in the energy density in Fig. \ref{dens1}.

%%%%%%%%%%%%%%%%%%%%%%%%%%%%%%%%%%%%%%%%%%%%%%%%%%%%%%%%%%%%%%%%%%
%%%%%%%%%%%%%%%%%%%%%%%%%%%%%%%%%%%%%%%%%%%%%%%%%%%%%%%%%%%%%%%%%%

\begin{figure}
	\includegraphics[{angle=0,width=8cm,height=6cm}]{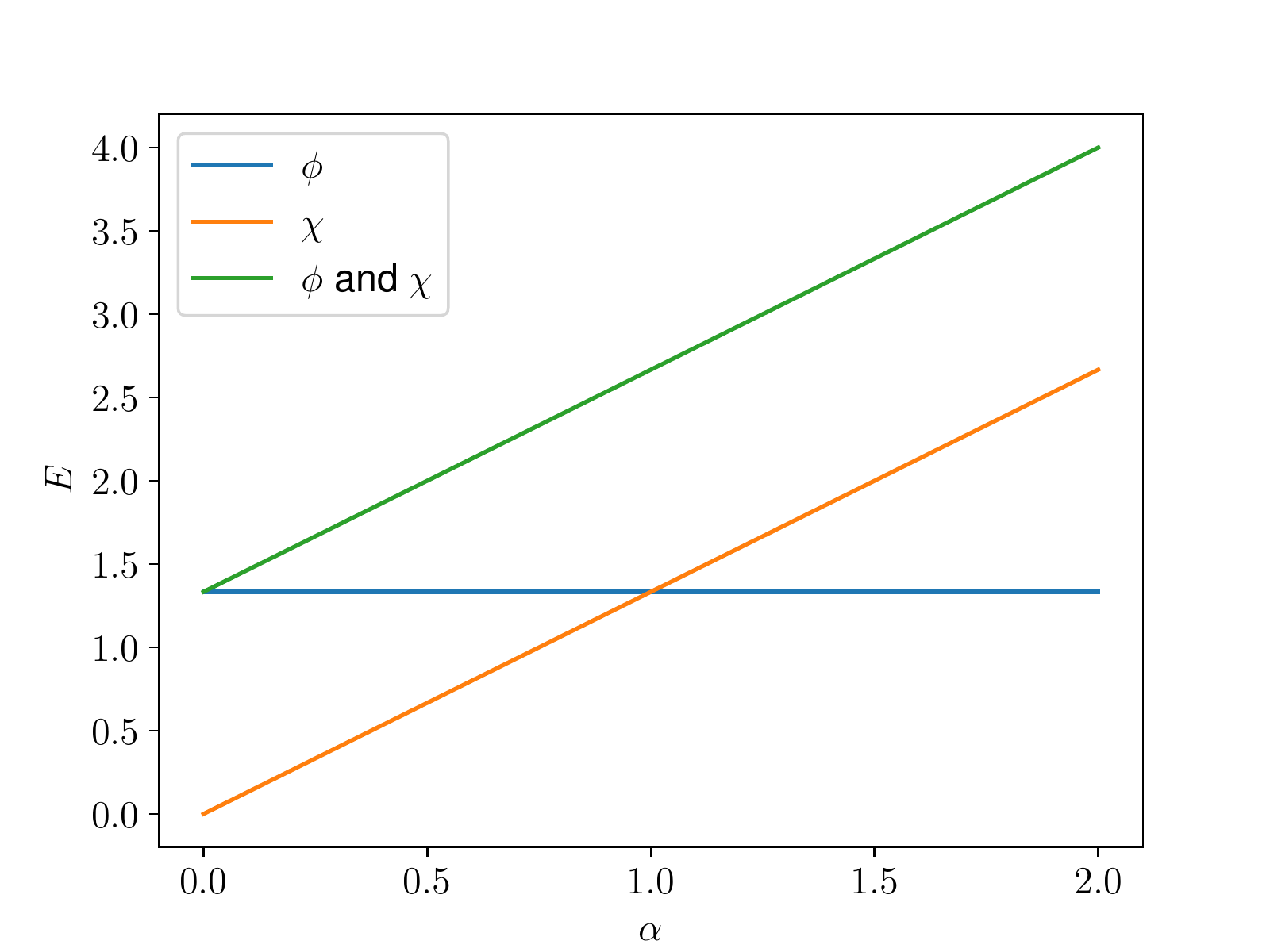}
	\caption{Energies or masses of the kinks as a function of $\alpha$.}
	\label{mass}
\end{figure}

The results presented are governed by geometrical constriction and the choice of the $f(\chi)$ modifies the center of the defect, however, it does not modify the tail of the solution. This enables the energy density to be integrated, resulting in $E_1=4/3$ and $E_2=4\alpha/3$. In Fig. \ref{mass}, we depict the behavior of energy (mass) as a function of $\alpha$. When $\alpha$ is small, the $\phi$ field is heavier than the $\chi$ field. This is altered for larger values of $\alpha$.

The above results were already discussed in Refs. \cite{blm,bbm}. The intention here was to make it clear that the parameters $\alpha$ and $\lambda$ directly contribute to changing the behavior of the solutions. In Ref. \cite{jab}, the geometrical constriction effect is observed in magnetic materials. In the present work, we want to investigate how the internal structure influences the kink-antikink collision process due to modifications in the geometrical constriction. As we can see, the kink-type configuration is a composed structure, governed by $\chi_K(x)$ in \eqref{chi} and $\phi_K(x)$ in \eqref{phi}. It is controlled by $\alpha$ and $\lambda$ and aggregates distinct forms and energy densities which deserve further investigation.

%%%%%%%%%%%%%%%%%%%%%%%%%%%%%%%%%%%%%%%%%%%%%%%%%%%%%%%%%%%%%%%%%%
%%%%%%%%%%%%%%%%%%%%%%%%%%%%%%%%%%%%%%%%%%%%%%%%%%%%%%%%%%%%%%%%%%

%%%%%%%%%%%%%%%%%%%%%%%%%%%%%%%%%%%%%%%%%%%%%%%%%%%%%%%%%%%%%%%%%%%%%

\section{Stability Analysis}
\label{sec:stab}
%%%%%%%%%%%%%%%%%%%%%%%%%%%%%%%%%%%%%%%%%%%%%%%%%%%%%%%%%%%%%%%%%%%%%

In this section we investigate the classical stability of the solutions. Small perturbations around the static solution are considered, therefore the fields can be represented as $\phi(x,t)=\phi(x)+\eta(x)\cos(\omega t)$ and $\chi(x,t)=\chi(x)+\psi(x)\cos(\omega t)$. The classical stability of solutions in a class of systems with two coupled scalar fields was explored in Refs. \cite{barisa,banari}. In Ref. \cite{adolroqwza}, stability analysis has been performed to determine the flow of normal modes. The authors showed that spectral walls are formed in models with multiple scalar fields with the presence of two zero modes. A similar model, in terms of multiple zero modes, is presented in Ref. \cite{nicoolewere}. 

Now, substituting $\phi(x,t)$ and $\chi(x,t)$ in the equations of motion \eqref{eqm1} and \eqref{eqm2}, we obtain, up to first order in $\eta$ and $\psi$
\begin{align}
    -f\eta_{xx}-f_\chi\chi_x\eta_x-f_\chi\phi_x\psi_x
    +V_{\phi\phi}\eta+\left(V_{\phi\chi}-\phi_x\chi_xf_{\chi\chi}-\phi_{xx}f_\chi\right)\psi=\omega^2f\eta,\\
    -\psi_{xx}+f_\chi\phi_x\eta_x
    +V_{\chi\phi}\eta+\left(V_{\chi\chi}+\frac{1}{2}f_{\chi\chi}\phi_x^2\right)\psi=\omega^2\psi.
\end{align} 
In general, these equations cannot be solved analytically. It is also a complicated task to find the spectrum numerically, due to the coupling between the fields. Luckily, there are a few special cases of interest where some analytical treatment is possible. We will first consider the solution described by Eqs.~\eqref{chi} and \eqref{phi}. In such case, it is not possible to compute the excited modes analytically, but the two zero modes are obtained as
\begin{align}
&\eta_0=\left(\frac{1+\lambda\tanh^2(\alpha x)}{1+\lambda}\right)\sech^2\left(y_0+x-\frac{\lambda}{\alpha(1+\lambda)}\tanh(\alpha x)\right),&\psi_0=\sech^2(\alpha x);\\
&\eta_0=\sech^2\left(y_0+x-\frac{\lambda}{\alpha(1+\lambda)}\tanh(\alpha x)\right),&\psi_0=0.
\end{align}
The first one is related to the translation of both fields simultaneously, and the second one is related to the translation of only the first component, $\phi$, in the direction of constant energy.

The second case of interest contains a trivial second component, $\chi(x)=\pm1$, and a kink in the first component $\phi(x)=\tanh(x)$. Then the stability equation becomes
\begin{align}
\label{eq:stab:eta2}
   -\eta_{xx}-f_\chi\phi_x\psi_x-2\phi_{xx}f_\chi\psi+(6\phi^2-2)\eta=\omega^2\eta,\\
    -\psi_{xx}+f_\chi\phi_x\eta_x
    -f_\chi\phi_{xx}\eta+\left[4\alpha^2+\left(f_{\chi}\right)^2(1-\phi^2)^2\right]\psi=\omega^2\psi.
\label{eq:stab:psi2}
\end{align}
Again it is not possible to compute the excited modes analytically, but the zero mode is obtained as
\begin{align}
    \eta_0=\sech^2(x),\quad\psi_0=0.
\end{align}
Here there is only one zero mode because one of the fields is in a trivial configuration.

\begin{figure}
	\includegraphics[{angle=0,width=15cm,height=6cm}]{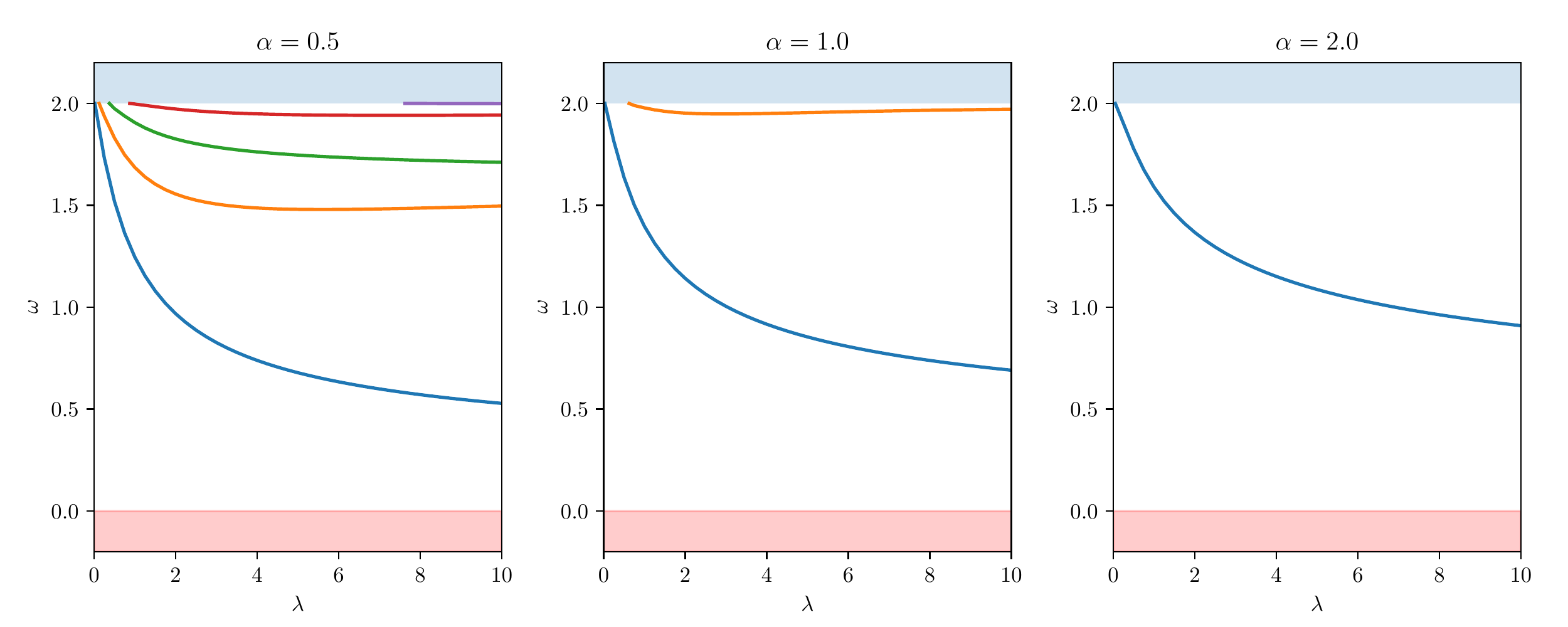}
	\caption{Excitation frequencies $\omega$ for different values of $\lambda$ and $\alpha$.}
	\label{boundstate}
\end{figure}

The third case of interest contains a trivial first component $\phi(x)=\pm1$ and a kink in the second component $\chi(x)=\tanh(\alpha x)$. In such case, we obtain the decoupled equations
\begin{align}
\label{eq:stab:eta3}
    -\frac{1}{f}\frac{d}{dx}(f\eta_{x})+\frac{4}{f^2}\eta=\omega^2\eta,\\
    -\psi_{xx}+\alpha^2(6\chi^2-2)\psi=\omega^2\psi.
\label{eq:stab:psi3}
\end{align}
The discrete spectrum coming from Eq.~\eqref{eq:stab:psi3} consists of a zero mode and a vibrational mode with frequency $\omega^2=3\alpha^2$. They are
\begin{align}
    \psi_0=\sech^2(\alpha x),\quad\psi_1=\tanh(\alpha x)\sech(\alpha x).
\end{align}
Moreover, Eq.~\eqref{eq:stab:eta3} may lead to additional discrete modes depending on the values of $\alpha$ and $\lambda$. We analyzed the behavior of the perturbation potential for values of $\alpha$ and $\lambda$. We realized that for $\alpha=0.5$, we have a wider potential. On the other hand, an increase in $\alpha$ indicates a thinner potential. This has consequences for obtaining vibrational states. In order to solve the Schr\"odinger-like equation numerically, we use the finite element method with quadratic approximation over a domain of $x=[-L,L]$, where $L=20$. In Fig. \ref{boundstate}, the discrete frequencies are illustrated as a function of $\lambda$ for $\alpha=0.5$, $1.0$, and $2.0$. For nonzero $\lambda$, a vibrational mode always exists. As observed, increasing $\lambda$ increases the number of vibrational states for $\lambda=0.5$ and $1.0$. Due to the extra number of bound states, the exchange of energy between the kinks becomes complicated, leading to a suppression of the two-bounce windows.

%%%%%%%%%%%%%%%%%%%%%%%%%%%%%%%%%%%%%%%%%%%%%%%%%%%%%%%%%%%%%%%%%%
%%%%%%%%%%%%%%%%%%%%%%%%%%%%%%%%%%%%%%%%%%%%%%%%%%%%%%%%%%%%%%%%%%

%%%%%%%%%%%%%%%%%%%%%%%%%%%%%%%%%%%%%%%%%%%%%%%%%%%%%%%%%%%%%%%%%%%%%

\section{Numerical results}\label{sec3}

%%%%%%%%%%%%%%%%%%%%%%%%%%%%%%%%%%%%%%%%%%%%%%%%%%%%%%%%%%%%%%%%%%%%%

Let us now describe the numerical results concerning the scattering process of kinks. First, we discuss the possible collision scenarios on general grounds. On each topological sector, it is possible to construct a kink-antikink collision. Due to symmetry, they are reduced to three cases. The first possibility is to scatter BPS solutions with constant $\chi$, and one may also scatter solutions with constant $\phi$. The final and more complicated scenario consists of scattering the BPS solution described by Eqs. \eqref{chi} and \eqref{phi}.

We will start our analysis with the more complicated scenario. In such a case, as there are two zero modes, the collision is specified by 8 constants, four positions $x_0^L,x_0^R,y_0^L,y_0^R$ and four velocities $\dot{x}_0^L,\dot{x}_0^R,\dot{y}_0^L,\dot{y}_0^R$, where $L$ ($R$) refers to the kink located on the left-hand (right-hand) side. After changing coordinates to the center of mass frame, these are reduced to 3 positions and 3 velocities. After fixing the relative position $2x_0$ and relative velocity $2v$, there are still 4 free parameters, which are highly impractical to be fully explored. Therefore, we will focus on the case where the kinks are coincident and co-moving in both fields because this is the case where the effects of the kinetic modification in our model are more evident. In fact, in most cases without such restriction, the kinks in both fields are far apart and the collisions will effectively be a sequence of isolated simple collisions with either constant $\phi$ and $\chi$, which will be considered subsequently (see Sec.~\ref{secC}).

We proceed by solving the two equations of motion (Eqs. (\ref{eqm1}) and (\ref{eqm2})) in a box in the interval $-100<x<+100$ with a space step $\Delta x=0.05$. The partial derivatives with respect to $x$ were approximated using the five-point stencil. The resulting set of equations was integrated using a fifth-order Runge-Kutta method with adaptive step size. Moreover, we considered periodic boundary conditions and we fixed $x=\pm x_0=\pm 10$ for the initial symmetric position of the pair. We used the following initial conditions for scattering
\begin{eqnarray}
\phi(x,0,x_0,v)&=&\phi_K(x+x_0,0,v) - \phi_K(x-x_0,0,-v)-1\\
\dot{\phi}(x,0, x_0, v)&=&\dot{\phi_K}(x+x_0,0,v) - \dot{\phi_K}(x-x_0,0,-v),
\end{eqnarray} 
and
\begin{eqnarray}
\chi(x,0, x_0, v)&=&\chi_K(x+x_0,0,v) - \chi_K(x-x_0,0,-v)-1\\
\dot{\chi}(x,0, x_0, v)&=&\dot{\chi_K}(x+x_0,0,v) - \dot{\chi_K}(x-x_0,0,-v),
\end{eqnarray} 
where $\phi_K(x,t,v)=\phi_{K}(\gamma(x-vt))$ and $\chi_K(x,t,v)=\chi_{K}(\gamma(x-vt))$ means a boost for the static solution with $\gamma=(1-v^2)^{-1/2}$. In the following, we will discuss our main results of the scattering of kinks with internal structure.

We vary the parameters $\alpha$, $\lambda$, and initial velocity $v_i$ as a starting point for the analysis of kink-antikink scattering. Particularly, the results are complex due to the high nonlinearity and the presence of geometric constraints. The following discussion will be divided into two parts, the first dealing with the kink-antikink collision for small values of $\lambda$, and the second covering the interaction for large $\lambda$ values. Then, we will inspect the collision scenarios with either trivial $\phi$ or $\chi$.

%%%%%%%%%%%%%%%%%%%%%%%%%%%%%%%%%%%%%%%%%%%%%%%%%%%%%%%%%%%%%%%%%%%%%

\subsection{Small values of $\lambda$}\label{secA}

%%%%%%%%%%%%%%%%%%%%%%%%%%%%%%%%%%%%%%%%%%%%%%%%%%%%%%%%%%%%%%%%%%%%%

\begin{figure}
    \includegraphics[{width=16cm,height=12cm}]{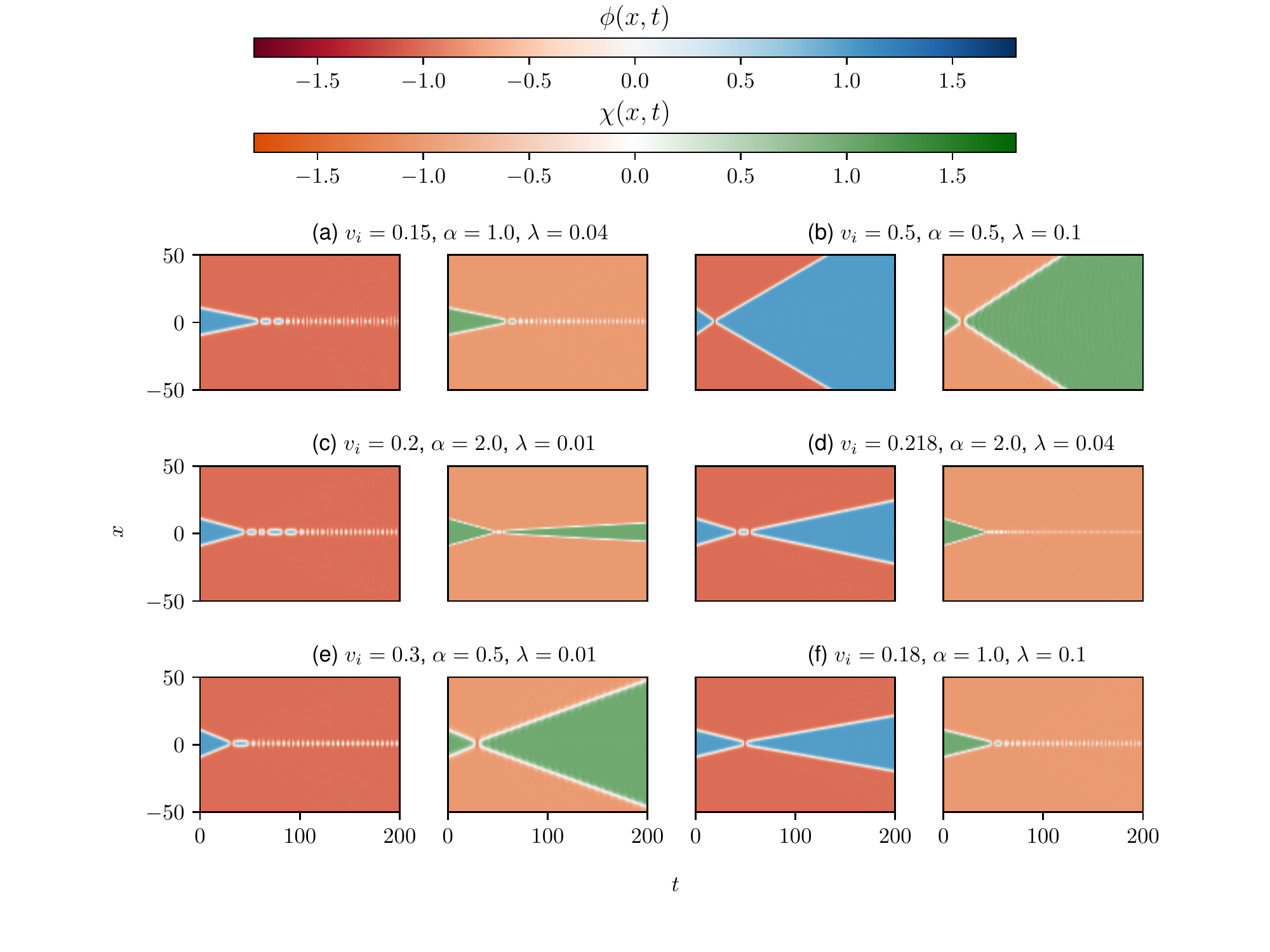}
    \caption{Evolution of the scalar fields in spacetime. Several values of the parameters are considered with small $\lambda$.}
	\label{cols1}
\end{figure}

Kink-antikink scattering for the case with internal structure and small values of $\lambda$ will be covered in this section. For $\lambda=0$, we know that the $\phi^4$ model is recovered. As a result, for this range of $\lambda$ values, the contribution of geometric constriction to the internal structure is lower. In the present case, the behavior of each field can be either annihilation with bion formation, reflection after a single bounce, and reflection after multiple bounces. As there are three possibilities for each field, there are nine possibilities in total. The collisions in $\phi$ and $\chi$ occur at the same time in addition to the two fields being coupled, allowing the kinks to exchange energy during the collision process. Several collisions with distinct parameters are shown in Fig. \ref{cols1}, illustrating the six most common cases. For instance, Fig. \ref{cols1}(c) depicts a collision with parameters $v_i=0.2$, $\alpha=2.0$ and $\lambda=0.01$. In such case, a bion is formed in the field $\phi$, i.e., the field oscillates erratically after the collision, emitting radiation. On the other hand, one also sees the field $\chi$ exhibits two-bounce scattering for the same parameters.

%%%%%%%%%%%%%%%%%%%%%%%%%%%%%%%%%%%%%%%%%%%%%%%%%%%%%%%%%%%%%%%%%%
%%%%%%%%%%%%%%%%%%%%%%%%%%%%%%%%%%%%%%%%%%%%%%%%%%%%%%%%%%%%%%%%%%

\begin{figure}
    \includegraphics[{width=16cm,height=12cm}]{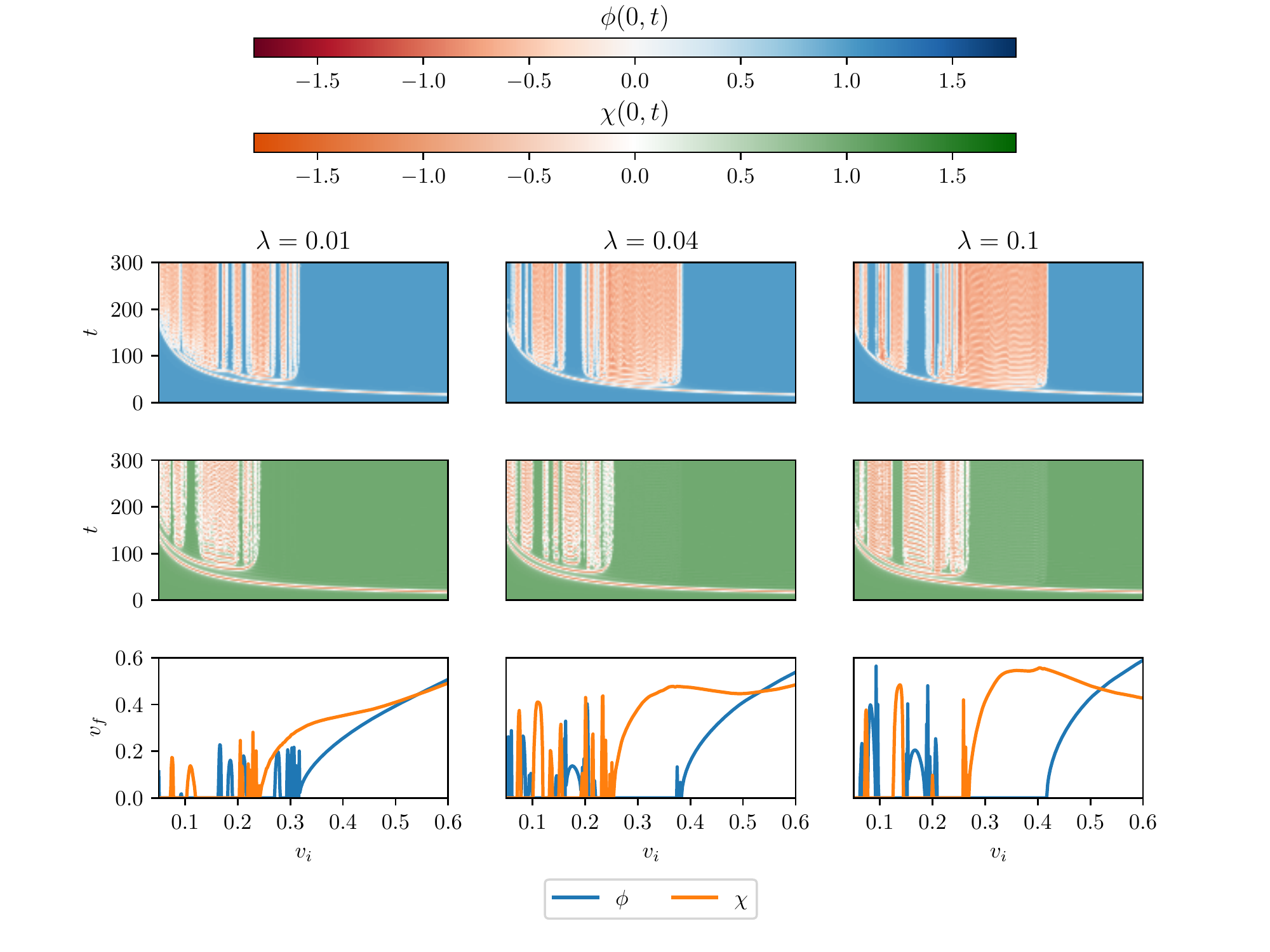}
    \caption{(First and second row) Evolution of scalar fields at the center of mass as a function of time and initial velocity. (Third row) Final velocity as a function of the initial one, using $\alpha=0.5$ and $\lambda=0.01$, $0.04$, $0.1$, from left to right.}
	\label{vel1}
\end{figure}

\begin{figure}
    \includegraphics[{width=16cm,height=12cm}]{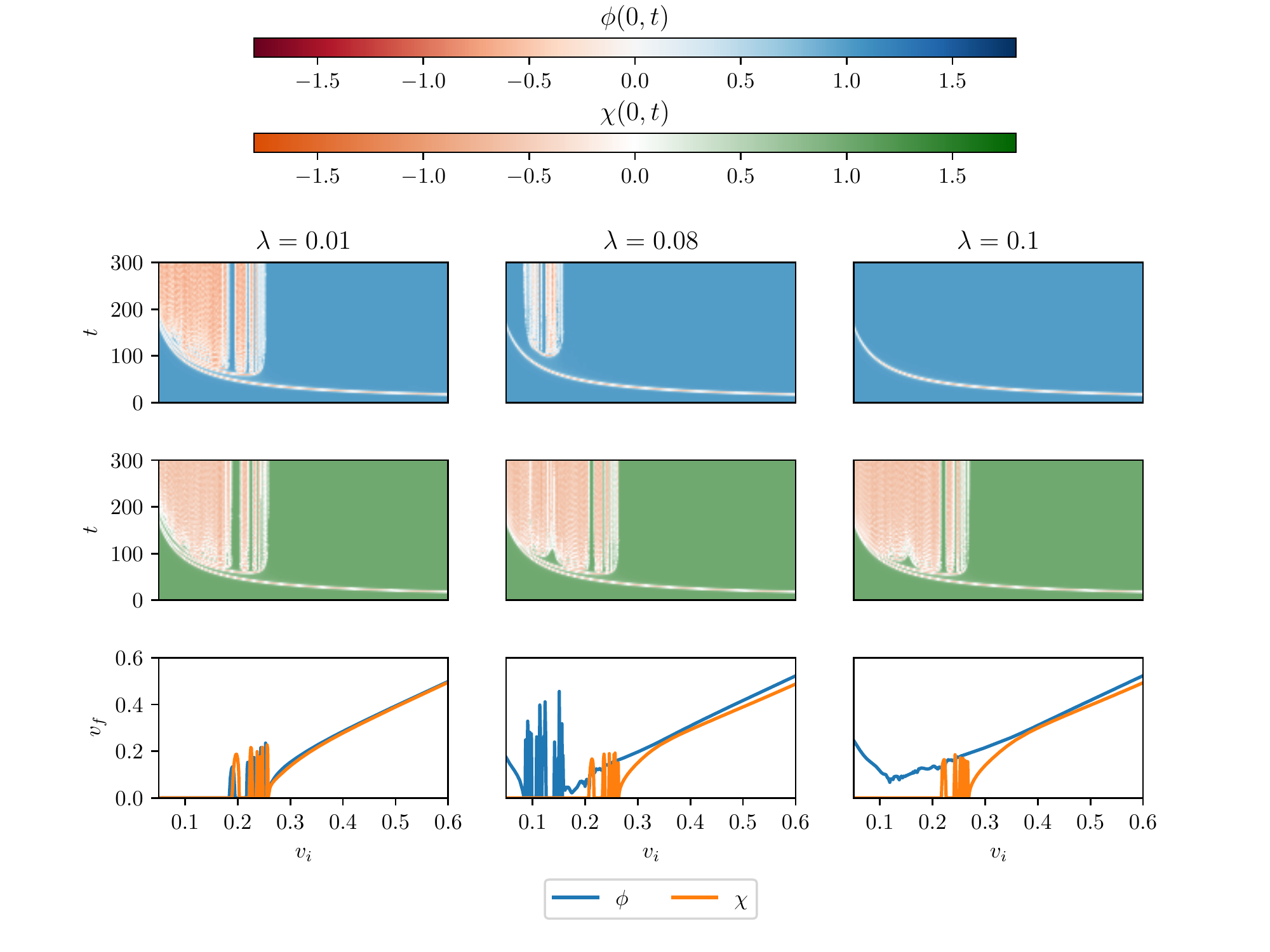}
    \caption{(First and second row) Evolution of scalar fields at the center of mass as a function of time and initial velocity. (Third row) Final velocity as a function of the initial one, using $\alpha=1.0$ and $\lambda=0.01$, $0.08$, $0.1$, from left to right.}
	\label{vel2}
\end{figure}

\begin{figure}
    \includegraphics[{width=16cm,height=12cm}]{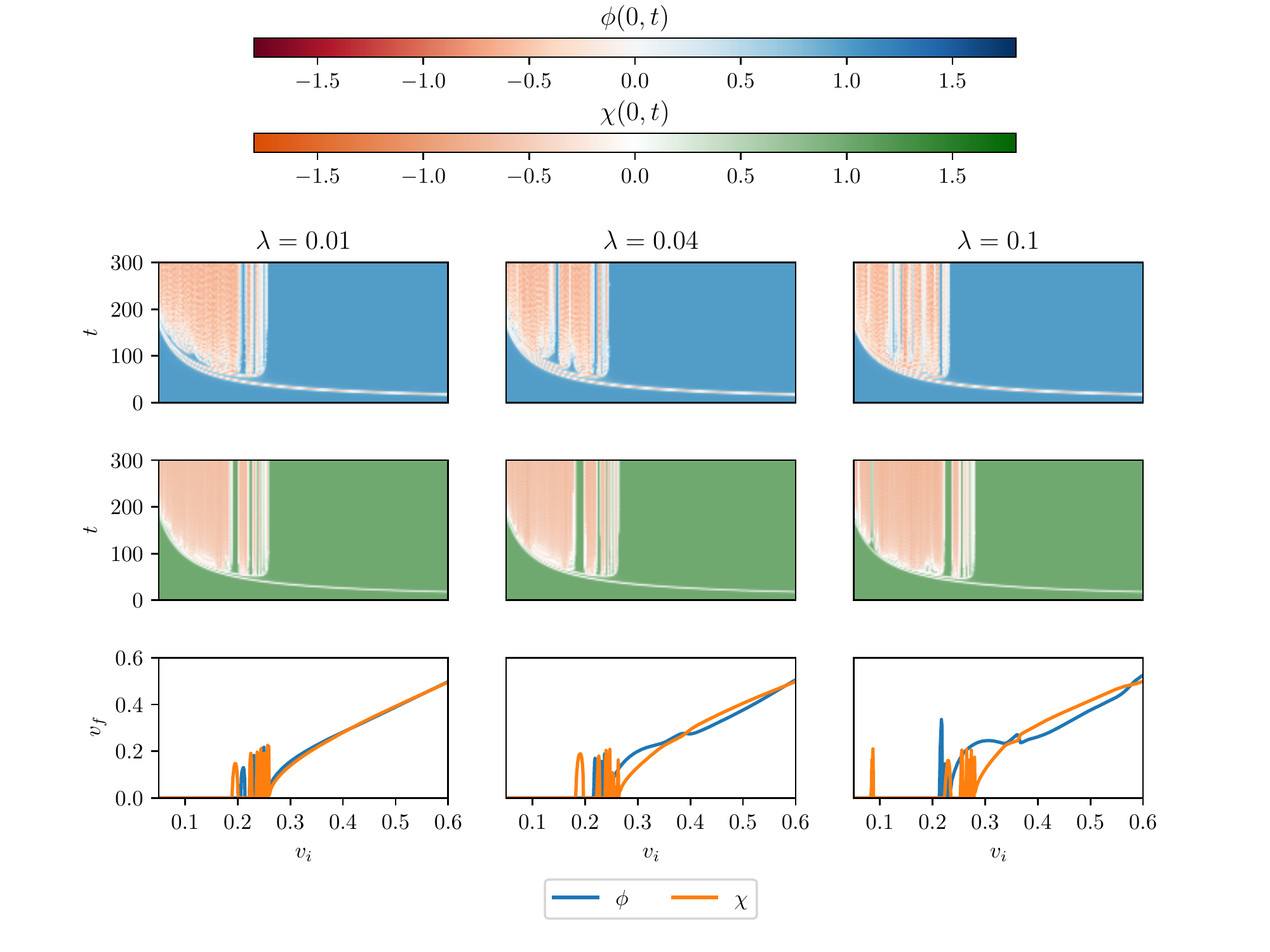}
    \caption{(First and second row) Evolution of scalar fields at the center of mass as a function of time and initial velocity. (Third row) Final velocity as a function of the initial one, using $\alpha=2.0$ and $\lambda=0.01$, $0.04$, $0.1$, from left to right.}
	\label{vel3}
\end{figure}

The structure of scattering for a small value of $\lambda$ is depicted in Figs. \ref{vel1}, \ref{vel2} and \ref{vel3} for $\alpha=0.5$, $1.0$, and $2.0$, respectively. The values of $\alpha$ were chosen in order to probe regimes where the contribution to the energy from the field $\chi$ is respectively smaller, equal, and larger than the one from the field $\phi$. In the figures, we show in each column, the values of the fields at the center of mass as a function of time and the initial velocity $v_i$, and we show the final velocity at the bottom. The colormaps can be interpreted as follows. A white line in the horizontal direction appears at every bounce, and blue (green) vertical lines appear when a kink is formed in the field $\phi$ ($\chi$). For easier visualization, the final velocities are only shown for the cases where the kinks reflect after one or two bounces.

%%%%%%%%%%%%%%%%%%%%%%%%%%%%%%%%%%%%%%%%%%%%%%%%%%%%%%%%%%%%%%%%%%
%%%%%%%%%%%%%%%%%%%%%%%%%%%%%%%%%%%%%%%%%%%%%%%%%%%%%%%%%%%%%%%%%%

%%%%%%%%%%%%%%%%%%%%%%%%%%%%%%%%%%%%%%%%%%%%%%%%%%%%%%%%%%%%%%%%%%
%%%%%%%%%%%%%%%%%%%%%%%%%%%%%%%%%%%%%%%%%%%%%%%%%%%%%%%%%%%%%%%%%%

%%%%%%%%%%%%%%%%%%%%%%%%%%%%%%%%%%%%%%%%%%%%%%%%%%%%%%%%%%%%%%%%%%
%%%%%%%%%%%%%%%%%%%%%%%%%%%%%%%%%%%%%%%%%%%%%%%%%%%%%%%%%%%%%%%%%%

When $\lambda=0.01$, we notice a resonant structure that is similar to the behavior of the $\phi^4$ model, as expected, but less so for $\alpha=0.5$. In particular, we observe that both the $\phi$ and $\chi$ fields have two-bounce windows for small velocities. For $\alpha=1.0$ and $\alpha=2.0$, the fields exhibit a slight shift in the location of the resonant structure compared to the $\phi^4$ model, but it is less pronounced for the field $\chi$. It indicates that $\chi$ influences more $\phi$ than the opposite case, which is consistent with the role of the $\chi$ in providing a constriction for $\phi$. Recall that a localized structure in the field $\chi$ is used to trap the other field and modify its configuration. There is one important detail, however. Although the boosted BPS solutions are being used as initial conditions, the system deviates from the BPS regime when the kinks superpose. One particular consequence is that the field $\chi$ cannot be considered to be completely independent of $\phi$ in the collision process.

Comparing Figs. \ref{vel1} and \ref{vel3}, we notice that the difference in mass between the fields plays an important role as the lighter kink is always the first to separate. This defines the first critical velocity, while the second one occurs when there is separation in both fields after the first bounce. It is clear from the figures that both critical velocities exist in all cases, that is, eventually the kinks separate as the initial velocity increases. Surprisingly, when the masses are equal the behavior is drastically altered. As $\lambda$ increases, a one-bounce window appears in $\phi$ for very small values of $v_i$ (see Fig. \ref{vel2} for $\lambda=0.08$), and eventually one observes that the kink separates in the field $\phi$ for all initial velocities. On the other hand, the behavior in $\chi$ is not significantly altered. So we see that the separation of kinks in the field $\phi$ and $\chi$ seem to occur independently. Such a result is expected because, as shown in Sec.~\ref{sec:stab}, the configuration where the kinks are coincident contains two zero modes, which means that there is a direction where they can be separated without energy cost.

The appearance of a fractal structure is clear from the colormaps and plot of the final velocities. In fact, the fractal structure can be even more complex in the present model because the resonance windows occur in both fields $\phi$ and $\chi$. Interestingly, in the center column of Fig.~\ref{vel1}, the windows alternate between the two fields, forming an intricate pattern. Moreover, as the first critical velocity is reached, the windows are strongly suppressed, meaning that it originates from the coupling between the fields.

Another interesting effect is that the final velocities may not be monotonically increasing after the critical values. The explanation comes one more time from the coupling between the fields $\phi$ and $\chi$. In Fig.~\ref{vel1}, we notice that when $\lambda=0.04$ and $0.1$ the final velocity of the field $\chi$ initially increases after the first critical velocity but then decreases when the second critical velocity is approached. This is expected from the fact that a great amount of energy is needed to separate the kink in the field $\phi$, which, for $\alpha=0.5$, is heavier compared to the kink in the field $\chi$. The decrease shows that some amount of energy is taken from the field $\chi$. Interestingly, the second critical velocity can be significantly larger than the first one. In Fig.~\ref{vel3} the final velocity is also not monotonic but in a different manner. The final velocity of the field $\phi$ has a sudden change when it becomes equal to the one of the field $\chi$. After the second critical velocity, both kinks are moving away from the center of mass, but one is more internal with respect to that point and the other is more external. When the two final velocities are equal, these two roles are interchanged.

Our careful analysis shows that many intricate structures are formed in the present model due to its complexity and nonlinear character. The behavior is very rich, even when the coupling parameter is small. Now, we turn to the more complicated scenario where the coupling between the two fields is strong.

%%%%%%%%%%%%%%%%%%%%%%%%%%%%%%%%%%%%%%%%%%%%%%%%%%%%%%%%%%%%%%%%%%%%%

\subsection{Large values of $\lambda$}\label{secB}

%%%%%%%%%%%%%%%%%%%%%%%%%%%%%%%%%%%%%%%%%%%%%%%%%%%%%%%%%%%%%%%%%%%%%

In this section, we will discuss kink-antikink scattering for the case with internal structure and large values of $\lambda$. In this range, the localized structure feels a stronger constriction. Thus, the initial profile of the field $\phi$ is composed of two subkinks due to the appearance of a plateau at the solution's center. The energy density structure is also affected, since the contribution from $f(\chi)$ leads to the formation of a central minimum in $\rho_1$ for larger values of $\lambda$.

\begin{figure}
\includegraphics[{width=16cm,height=12cm}]{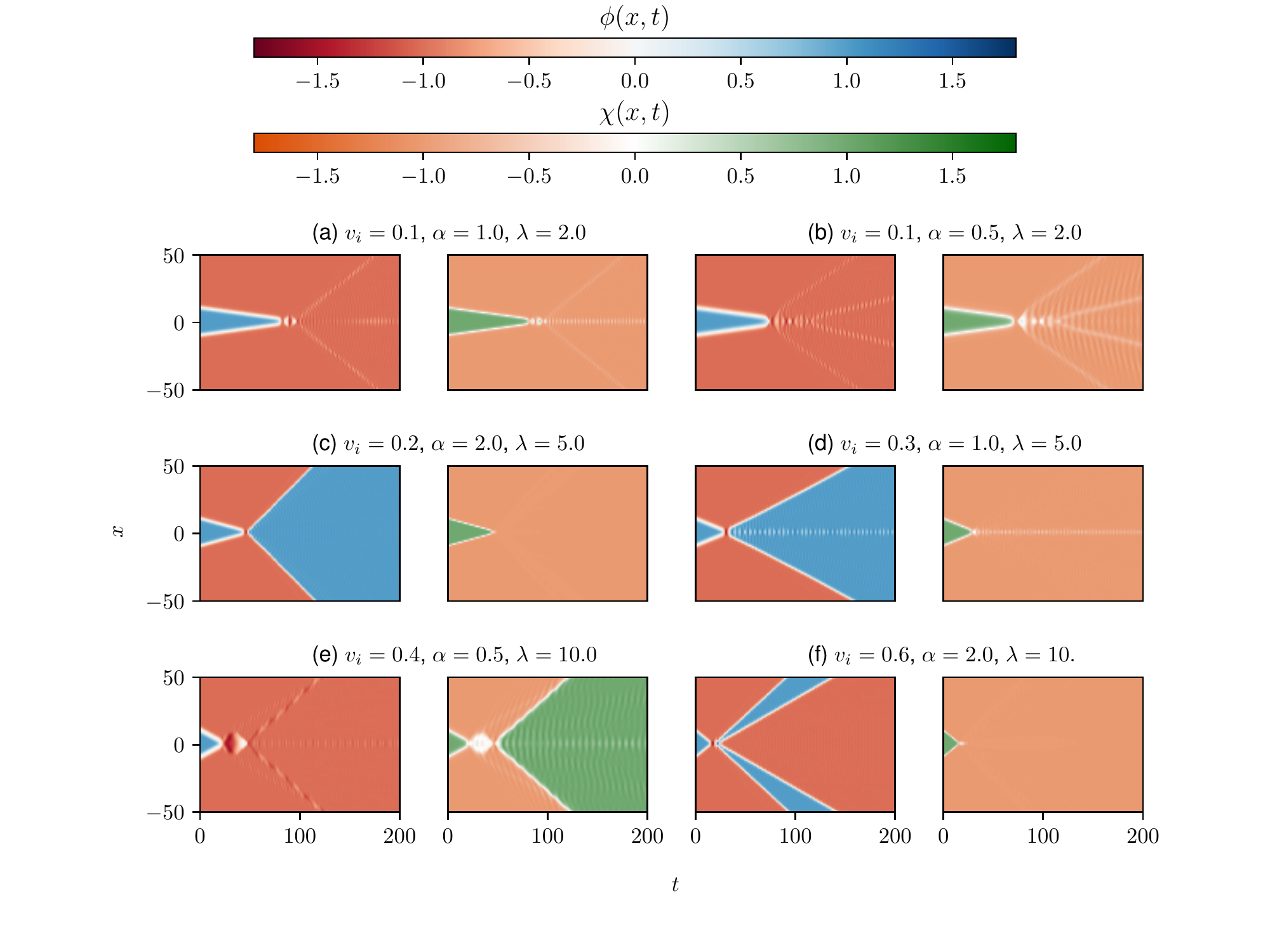}
    \caption{Evolution of the scalar fields in spacetime. Several values of the parameters are considered with large $\lambda$.}
	\label{cols2}
\end{figure}

In general, the scenarios reported above also occur for large values of $\lambda$. In addition, other interesting scenarios are observed. In fig.~\ref{cols2}, we illustrate a few examples. In panels (a) and (b) annihilation into three and four oscillons are shown, respectively. We use the term oscillon to designate long-lived oscillating pulses. In general, a strong constriction favors the appearance of multiple oscillons. In panels (c), two kinks are formed in the field $\phi$ after the collision, while the $\chi$ field is almost completely annihilated. In panel (d), two kinks are also formed in the field $\phi$ after the collision and a bion is formed in the other component. Interestingly, the bion also affects the field $\phi$ due to the strong coupling between the two. An escape scenario of kinks in the field $\phi$ is shown in panel (e), but it is not possible to count the number of bounces due to the highly nonlinear character of the model. The presence of kinks in $\chi$ has a visible effect on $\phi$. Finally, in panel (f), the scattering output is two pairs of kinks in the field $\phi$ and almost complete annihilation in the other component.

%%%%%%%%%%%%%%%%%%%%%%%%%%%%%%%%%%%%%%%%%%%%%%%%%%%%%%%%%%%%%%%%%%
%%%%%%%%%%%%%%%%%%%%%%%%%%%%%%%%%%%%%%%%%%%%%%%%%%%%%%%%%%%%%%%%%%

\begin{figure}
\includegraphics[{angle=0,width=16cm,height=9.6cm}]{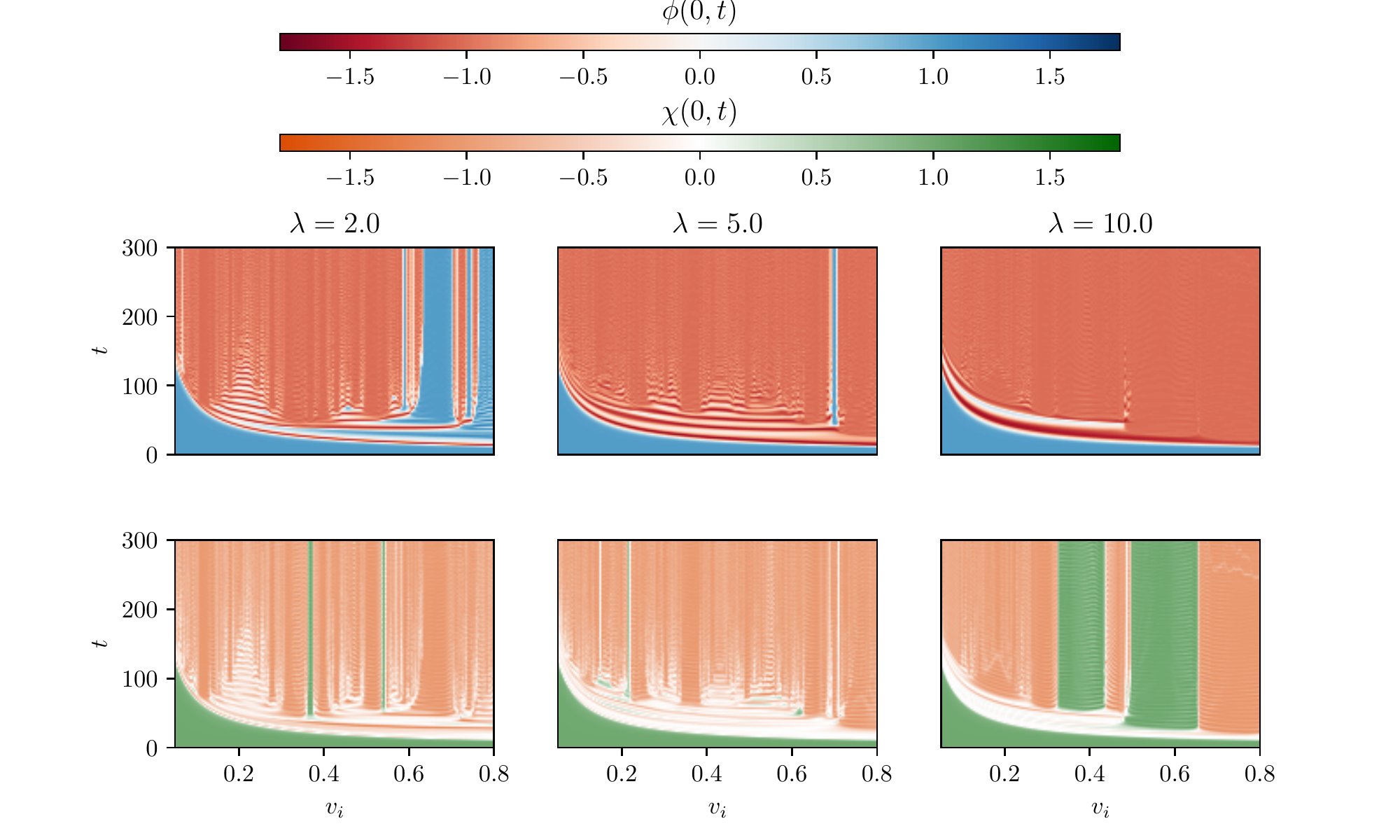}
    \caption{Evolution of scalar fields at the center of mass as a function of time and initial velocity. We set $\alpha=0.5$ and $\lambda=2.0$, $5.0$, $10.0$.}
	\label{center1}
\end{figure}

The structure of the scattering process for $\alpha=0.5$ is presented in Fig. \ref{center1}. It shows the value of the fields at the collision center, $\phi(0,t)$ and $\chi(0,t)$, as a function of time $t$ and initial velocity $v_i$. It is clear from the figure, that the system is drastically modified as $\lambda$ increases. For $\lambda=2.0$, the first critical velocity occurs for the field $\phi$, even though it does not contain the lightest kink. However, the velocity is very large and the second critical one was not observed in $\chi$. Interestingly, a structure of higher-bounce windows is still visible in the field $\phi$ near the first critical velocity. However, if $\lambda$ is increased further, neither the critical velocity nor the windows are observed anymore for the component $\phi$. So, in essence, the only outcome is kink-antikink annihilation. On the other hand, a few isolated escape windows are observed in $\chi$ in the three cases shown. For $\lambda=10.0$, they are quite large, with the smaller one on the left corresponding to the example given in Fig.~\ref{cols2}(e). Therefore, for large values of $\lambda$, critical velocities become ultrarelativistic, and complex escape patterns are not observed. Loosely speaking, it is possible to say that the system is too far from integrability for large values of $\lambda$.

%%%%%%%%%%%%%%%%%%%%%%%%%%%%%%%%%%%%%%%%%%%%%%%%%%%%%%%%%%%%%%%%%%
%%%%%%%%%%%%%%%%%%%%%%%%%%%%%%%%%%%%%%%%%%%%%%%%%%%%%%%%%%%%%%%%%%

\begin{figure}
\includegraphics[{angle=0,width=16cm,height=9.6cm}]{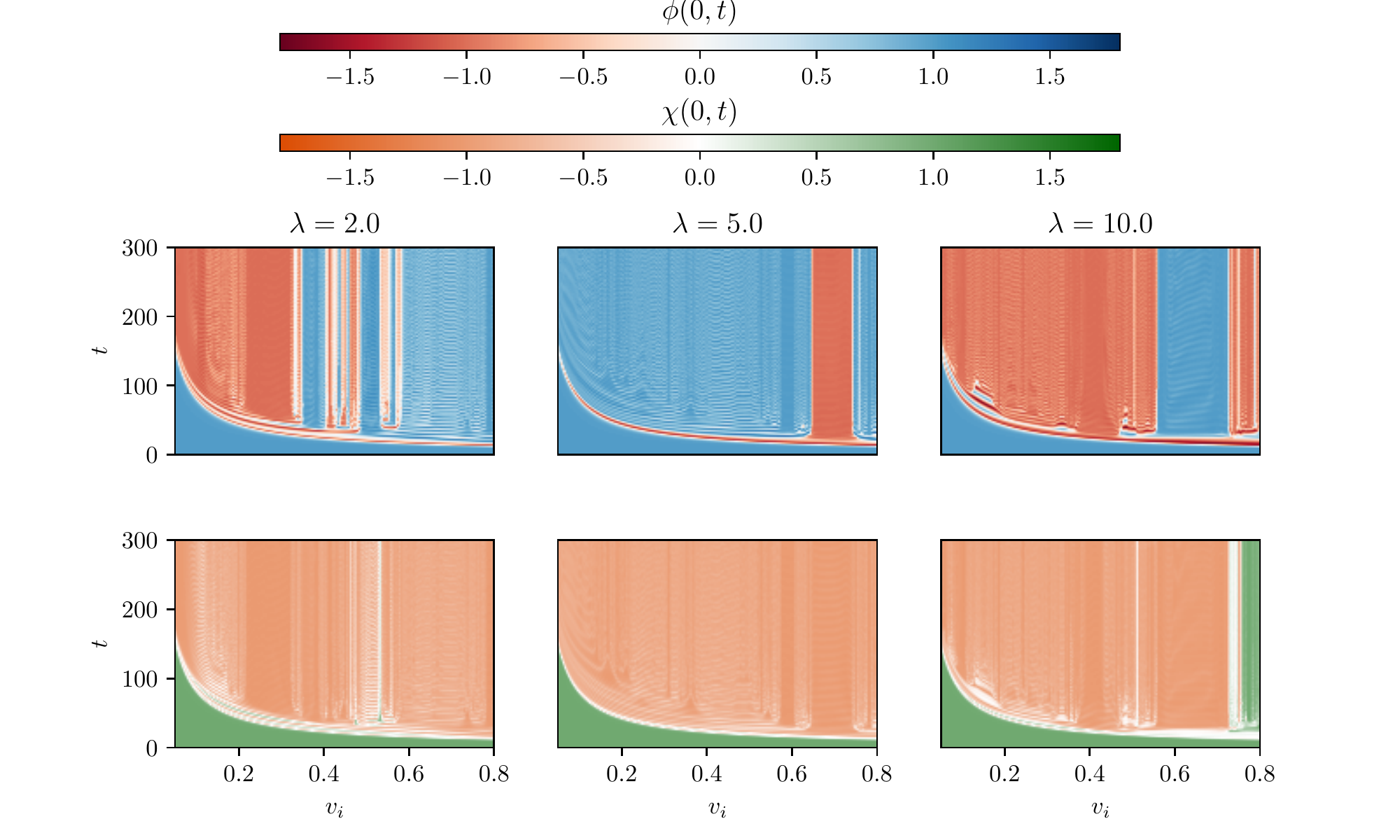}
    \caption{Evolution of scalar fields at the center of mass as a function of time and initial velocity. We set $\alpha=1.0$ and $\lambda=2.0$, $5.0$, $10.0$.}
	\label{center2}
\end{figure}

Let us now examine the scenario with $\alpha=1.0$. In Fig.~\ref{center2}, scattering outputs are shown for $\lambda=2.0$, $5.0$, and $10.0$. Again the system is drastically altered as $\lambda$ increases. For $\lambda=2.0$, there is a strong similarity between the outputs for $\alpha=0.5$ and $1.0$. This is quite surprising, given that the scattering outputs were quite distinct for small values of $\lambda$. One can see that there is annihilation in the component $\chi$ for most initial velocities in all cases. For $\lambda=5.0$, the behavior changes considerably as the kinks always escape in the field $\phi$ after a single bounce, and, increasing the velocity, there is a red region where an extra pair of kinks are formed in the same component. The formation of extra pairs is illustrated in Fig.~\ref{cols2}(f). As discussed in Refs. \cite{sifrgono,bazeia2023}, the difference in mass between incoming and outcoming kinks may result in oscillating pulses and extra kink-antikink pairs, such as the ones observed here. Similar results were observed in Ref. \cite{romanshnir}, where the authors obtained the formation of kink pairs from particle-like states via the excitation of oscillons. Moving on to the case with $\lambda=10.0$, the scattering output changes abruptly once more, illustrating the high sensitivity of the present model on its parameters. The results show mostly annihilation and two escape regions, one in $\phi$ and one in $\chi$.

%%%%%%%%%%%%%%%%%%%%%%%%%%%%%%%%%%%%%%%%%%%%%%%%%%%%%%%%%%%%%%%%%%
%%%%%%%%%%%%%%%%%%%%%%%%%%%%%%%%%%%%%%%%%%%%%%%%%%%%%%%%%%%%%%%%%%

\begin{figure}
\includegraphics[{angle=0,width=16cm,height=9.6cm}]{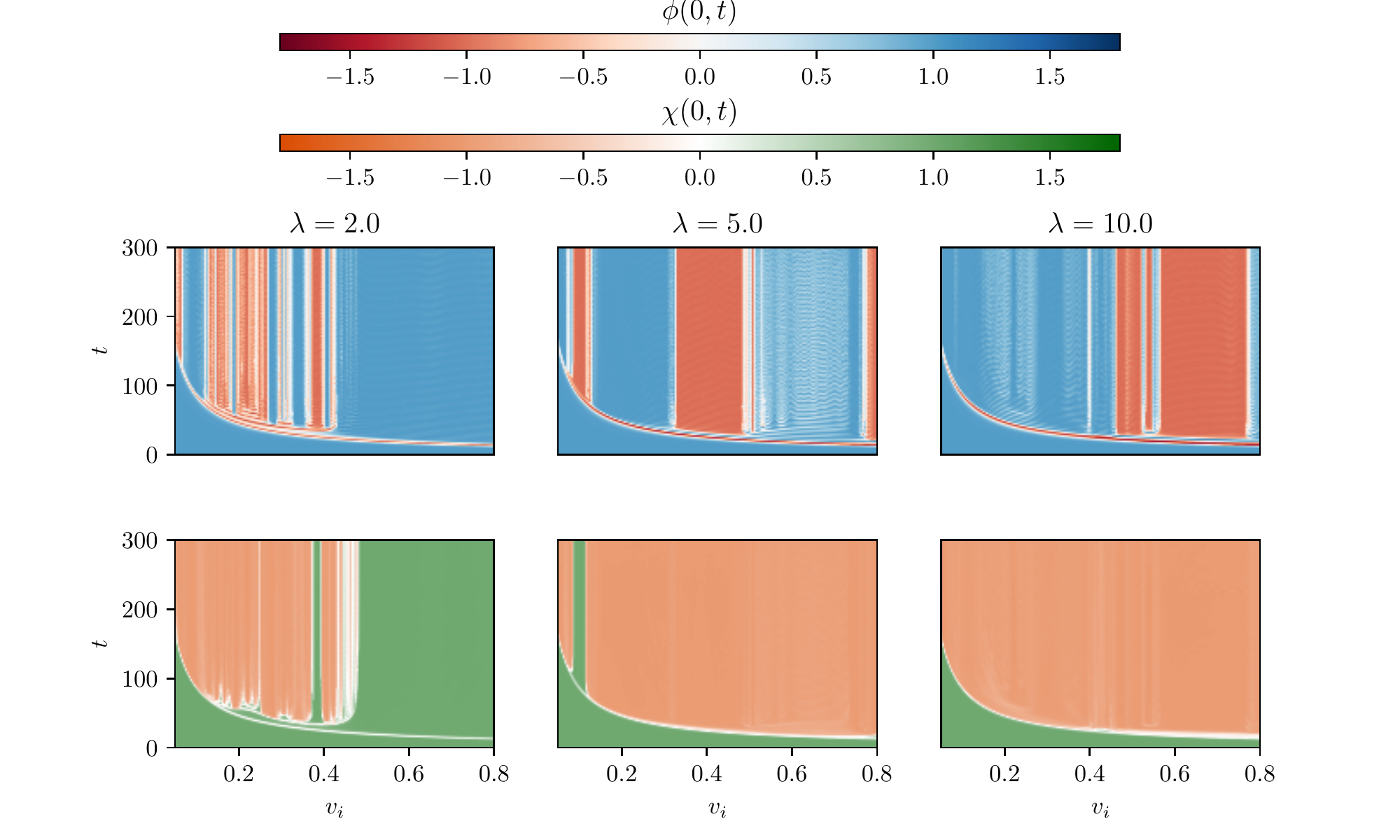}
    \caption{Evolution of scalar fields at the center of mass as a function of time and initial velocity. We set $\alpha=2.0$ and $\lambda=2.0$, $5.0$, $10.0$.}
	\label{center3}
\end{figure}

For $\alpha=2.0$, the scattering output is shown in Fig.~\ref{center3}. For $\lambda=2.0$, the observed behavior is still similar to the one for small $\lambda$. This indicates that the model is more robust for larger values of $\alpha$. The reason for this result is that the constricted field $\phi$ has a smaller back-reaction on the constricting field $\chi$ when the latter becomes more massive. However, the field $\chi$ eventually exhibits only annihilation if $\lambda$ is further increased. Interestingly, the $\phi$ field exhibits a rich pattern of one- and higher-bounce escape regions and even the formation of two kink-antikink pairs, occurring in the large red region with $\lambda=10.0$. The appearance of one-bounce windows is an intriguing behavior of the present model. They are allowed here because the initial kinks are more massive than the final ones. Therefore, the difference in mass can be enough to allow the final kinks to escape. The existence or absence of a one-bounce window depends on the complicated manner that the energy is redistributed after the interaction. In some cases, they are also linked with the almost complete annihilation occurring in the opposing field. Likewise, neither the second critical velocity nor the complex outcome patterns were observed in the present case. 

%%%%%%%%%%%%%%%%%%%%%%%%%%%%%%%%%%%%%%%%%%%%%%%%%%%%%%%%%%%%%%%%%%
%%%%%%%%%%%%%%%%%%%%%%%%%%%%%%%%%%%%%%%%%%%%%%%%%%%%%%%%%%%%%%%%%%

The lack of an ordered scattering output could also be linked to the excitation spectra of the kink solutions. For instance, we saw in Sec.~\ref{sec:stab} that some kinks possess more than a single vibrational mode. Such property is expected to suppress the appearance of a sequence of ordered resonance windows in most cases. This is a possible explanation for some of the behavior observed above.

After describing our results, a few comparisons are in order. The localized structures discussed in Ref. \cite{meoli,meoli1} shares similarities with the kink profile appearing here, also showing rich and intricate behavior. Interestingly, our findings also share similarities with those presented recently in Refs. \cite{maevvaza, mavacha}, which considered kink-antikink scattering coupled to an additional quantum field. The additional field favors the annihilation of the kinks in the original model, a tendency also observed here for large values of $\lambda$. 
Moreover, we observe that the model under consideration shows a variety of behaviors for large values of $\lambda$. We found that the formation of several oscillons is favored. They correspond to localized oscillating configurations that radiate slowly, and are prominent in bubble collisions \cite{cogleimu,zhaamcosa}. In this range of parameters, the escape patterns are less complex in general. For large values of $\alpha$, the escape pattern of the field $\phi$ is still intricate, but, in general, only annihilation is observed for the $\chi$ field. 

\subsection{Other kink-antikink collisions}\label{secC}

The two remaining kink-antikink collisions are the ones with either trivial $\phi$ or trivial $\chi$. The two cases are not equivalent due to the different roles of $\phi$ and $\chi$ in the Lagrangian. First, notice that setting $\phi=\pm 1$ solves Eq.~\eqref{eqm1}, and Eq.~\eqref{eqm2} reduces to the $\chi^4$ model. Therefore, the behavior is already well-known. In short, the output alternates between resonance windows and annihilation with bion formation for velocities below a critical one $v_c\simeq0.26$. Then, for larger velocities, the kinks reflect. For more details, we refer the reader to Ref.~\cite{camp1}.

%%%%%%%%%%%%%%%%%%%%%%%%%%%%%%%%%%%%%%%%%%%%%%%%%%%%%%%%%%%%%%%%%%
\begin{figure}
\includegraphics[{angle=0,width=16cm,height=9.6cm}]{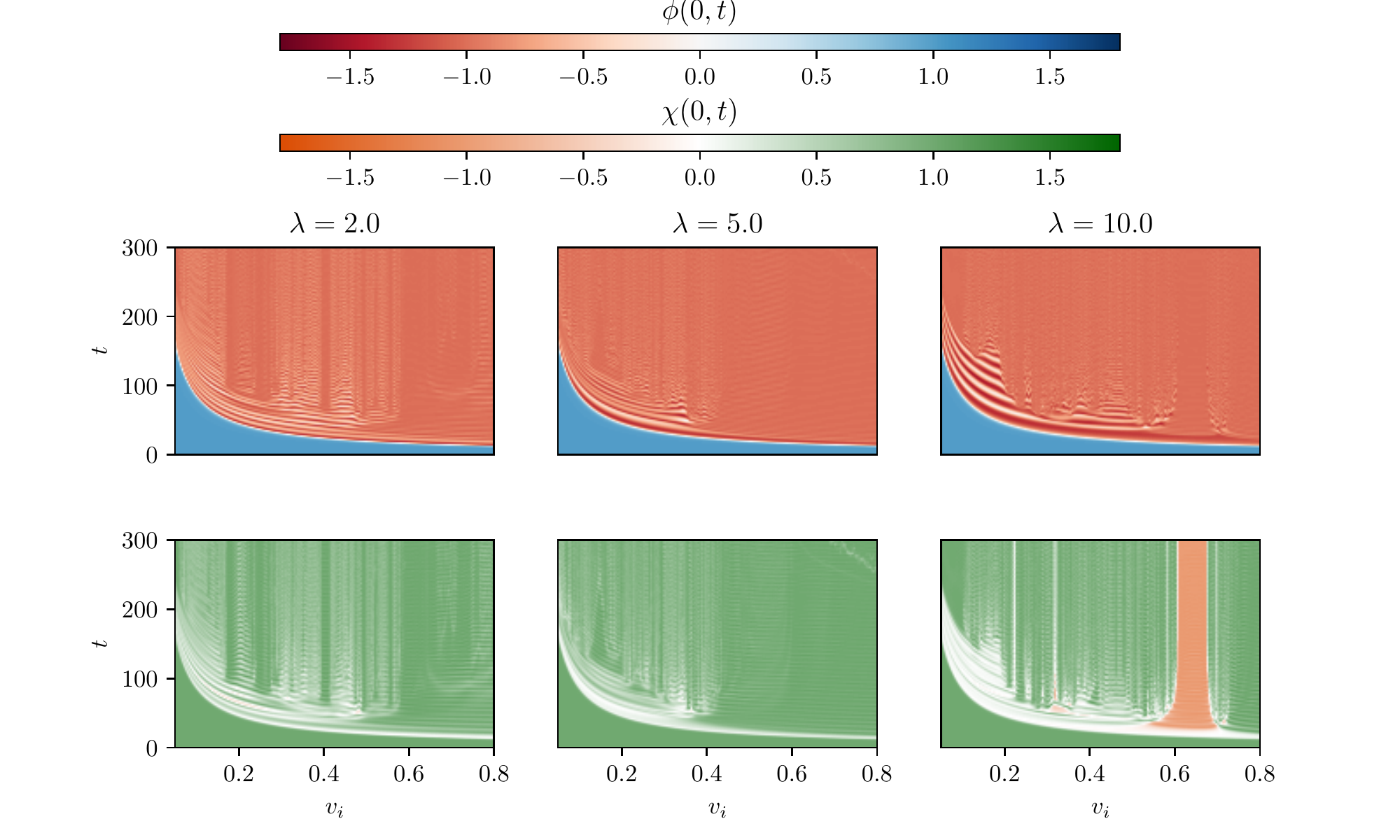}
    \caption{Evolution of scalar fields at the center of mass as a function of time and initial velocity. The initial condition is composed of kinks with trivial $\chi$. We set $\alpha=0.5$ and $\lambda=2.0$, $5.0$, $10.0$.}
	\label{center4}
\end{figure}
%%%%%%%%%%%%%%%%%%%%%%%%%%%%%%%%%%%%%%%%%%%%%%%%%%%%%%%%%%%%%

The last case is the one with trivial $\chi$. Setting $\chi=\pm 1$ does not solve Eq.~\eqref{eqm2} in general. Therefore, the fields do not decouple. The result for $\alpha=0.5$ and $\lambda=2.0$, $5.0$, and $10.0$ is shown in Fig.~\ref{center4}. After the first bounce, bions are formed in both fields. For this reason, it is very difficult for the kinks to separate. Increasing $\lambda$, it becomes more likely to transfer energy to the field $\chi$. Thus, there exists one escape region in $\chi$, which contains the lightest kink, occurring for $\lambda=10.0$ and $v_i\simeq0.65$.

%%%%%%%%%%%%%%%%%%%%%%%%%%%%%%%%%%%%%%%%%%%%%%%%%%%%%%%%%%%%%
\begin{figure}
\includegraphics[{angle=0,width=16cm,height=9.6cm}]{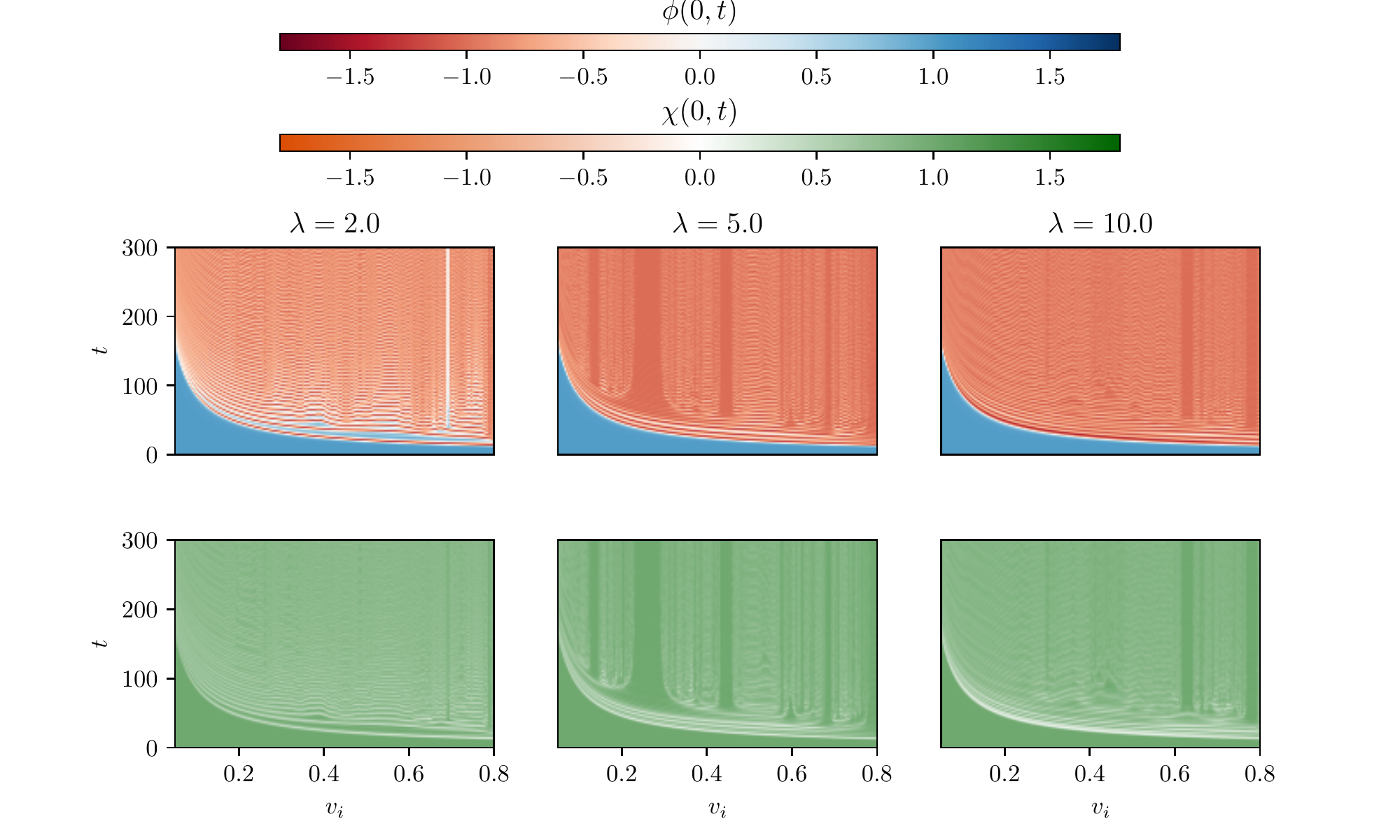}
    \caption{ Evolution of scalar fields at the center of mass as a function of time and initial velocity. The initial condition is composed of kinks with trivial $\chi$. We set $\alpha=1.0$ and $\lambda=2.0$, $5.0$, $10.0$.}
	\label{center5}
\end{figure}
%%%%%%%%%%%%%%%%%%%%%%%%%%%%%%%%%%%%%%%%%%%%%%%%%%%%%%%%%%%%%

The scattering output for $\alpha=1.0$ is shown in Fig.~\ref{center5}. The formation of bions in both fields hinders the separation of the kinks, similar to the previous case. However, the kinks in $\phi$ and in $\chi$ have the same mass now, and no escape windows are observed. One key difference between the results in the current section and the previous one is that the mass contained in the initial configuration is smaller because only one field component is nontrivial. Accordingly, the separation of kinks becomes more difficult.

%%%%%%%%%%%%%%%%%%%%%%%%%%%%%%%%%%%%%%%%%%%%%%%%%%%%%%%%%%%%%%
\begin{figure}
\includegraphics[{angle=0,width=16cm,height=9.6cm}]{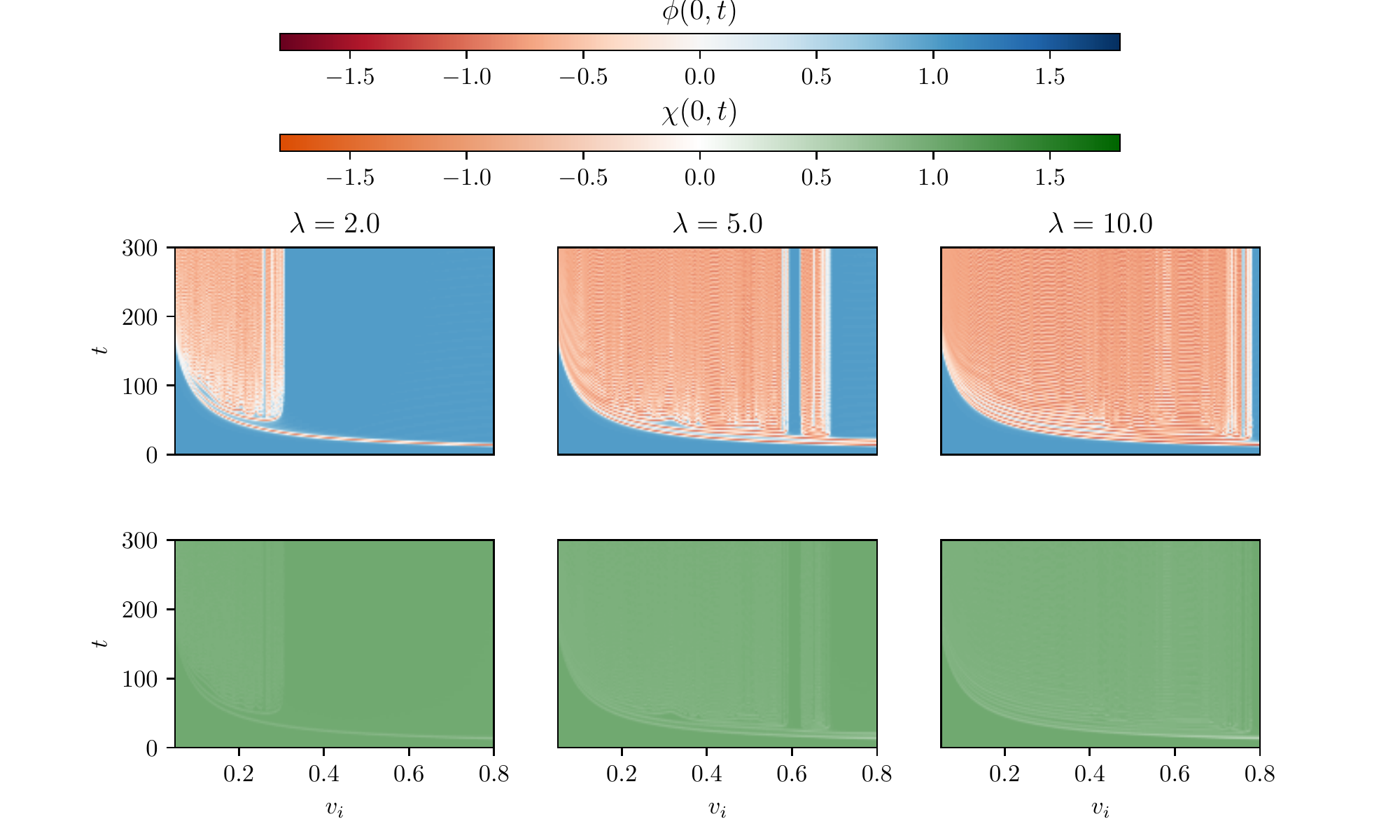}
    \caption{Evolution of scalar fields at the center of mass as a function of time and initial velocity. The initial condition is composed of kinks with trivial $\chi$. We set $\alpha=2.0$ and $\lambda=2.0$, $5.0$, $10.0$.}
	\label{center6}
\end{figure}
%%%%%%%%%%%%%%%%%%%%%%%%%%%%%%%%%%%%%%%%%%%%%%%%%%%%%%%%%%%%%%%%%%

Finally, the case with $\alpha=2.0$ is presented in Fig.~\ref{center6}. Now, the kink in $\chi$ is more massive than the initial ones. Similarly to Fig.~\ref{center3}, the field $\phi$ seems to be less affected by the other component, as the result for $\lambda=2.0$ is not far from to $\phi^4$ model. However, very few resonance windows are observed, which could be linked to the stability equation of the two components being coupled and differing from the pure $\phi^4$ theory. For $\lambda=5.0$ and $10.0$, the field $\chi$ is even more likely to be excited. Therefore, the separation of the kinks is hindered again.

%%%%%%%%%%%%%%%%%%%%%%%%%%%%%%%%%%%%%%%%%%%%%%%%%%%%%%%%%%%%%%%%%%%%%%%%%%

\section { Conclusions } \label{sec4}

%%%%%%%%%%%%%%%%%%%%%%%%%%%%%%%%%%%%%%%%%%%%%%%%%%%%%%%%%%%%%%%%%%%%%%%%%%

The purpose of this work was to investigate the kink scattering process in models with two scalar fields in the presence of geometric constrictions. This model is characterized by the introduction of an auxiliary function that modifies the kinematic part associated with the other scalar field. Such a change in the Lagrangian generates interesting results on the internal structure of the solution, such as the appearance of a two-kink structure, similar to the case of magnetic domain walls \cite{jab}. Another possibility regarding the use of geometric constrictions is that the function $f(\chi)$  can also be considered to control the internal structure in brane scenarios \cite{Ferreira}. It was demonstrated in Ref. \cite{blm} that the function $f(\chi)$ simulates the presence of a geometric constriction; however, even though it appears in the equation of motion, such function does not contribute to the BPS solution's total energy, which is solely dependent on the function $W(\phi,\chi)$. In this case, the scalar field $\chi$ can be resolved independently and gives rise to a kink that interferes with the configuration of the field $\phi$. The choice of $f(\chi)=\frac{1+\lambda}{1+\lambda\chi^2}$ has a direct impact on the internal structure of this field. 

We developed the linear perturbation study for the cases of interest. However, due to the interaction between the fields, the stability equations are typically too complicated to yield any analytical or even numerical solutions. As a result, we mainly highlight the presence of the two zero modes in the scenario with internal structure and also the increase in the number of bound states with the increase of $\lambda$ in the scenario with trivial $\phi$. The study of kink collisions within geometrically constrained systems revealed rich and intricate dynamics. In our numerical analysis, we observed how the presence of internal structure can influence the propagation and formation of kinks. Importantly, the fields collide simultaneously, allowing the kinks to exchange energy during the dynamics.

As a first example, we develop kink-antikink scattering for small values of $\lambda$. In that case, the outcomes demonstrated that the fields can annihilate, reflect after a single bounce, or reflect after multiple bounces. We selected $\alpha$ values to examine how massive solutions affect the collision process. The difference in mass between the fields plays an important role as the lighter kink is always the first to separate. Moreover, rich resonant patterns are observed, including a window structure that alternates between the two fields. In many cases, the behavior in the component $\phi$ deviates more from the isolated $\phi^4$ theory, compared to the component $\chi$. This is related to the influence of the fields, where the $\chi$ is responsible for promoting a constriction in $\phi$.

In the region for higher $\lambda$ values, the localized structure feels the geometric deformation more strongly. The kink interaction process is richer and more intricate due to the higher coupling between the fields. In particular, our results for both fields include oscillating pulses formation, as well as an alternating behavior between escape and annihilation with bion formation, forming a fractal structure. One explanation for the change in resonant windows is related to the distribution of the large energy coming from the initial kink configuration. 

Finally, we examined collisions of kinks with either trivial $\phi$ or $\chi$. In the first case, the well-known $\chi^4$ model is recovered. In the second one, kink annihilation is enhanced due to the decrease in the mass of the initial kinks, combined with the possibility of the collision in $\phi$ exciting the field $\chi$ and losing energy. The results obtained in this work contribute to the theoretical framework of field theory and still have the potential to inspire advances at the nanometric scale.

%%%%%%%%%%%%%%%%%%%%%%%%%%%%%%%%%%%%%%%%%%%%%%%%%%%%%%%%%%%%%%%%%%%%%%%%%%%%%

\section{Acknowledgements}
 This study was financed in part by Conselho Nacional de Desenvolvimento Cient\'ifico e Tecnol\'ogico, Grants No. 303469/2019-6 (DB) and No. 150166/2022-2 (JGFC), Funda\c c\~ao de Amparo \`a Pesquisa e ao Desenvolvimento do Maranh\~ao, Grant No. 00920/19 (FCS), Funda\c{c}\~ao de Amparo \`a Ci\^encia e Tecnologia de Pernambuco, Grant No. BFP-0013-1.05/23 (JGFC). This study was financed in part by the Coordena\c c\~ao
de Aperfei\c coamento de Pessoal de N\'ivel Superior - Brasil (CAPES) - Finance Code 001. Paraiba State Research Foundation, Grant No. 0015/2019 (DB). The simulations presented here were performed in the supercomputer SDumont of the Brazilian Laboratory LNCC (Laborat\'orio Nacional de Computa\c{c}\~ao Cient\'ifica).

%%%%%%%%%%%%%%%%%%%%%%%%%%%%%%%%%%%%%%%%%%%%%%%%%%%%%%%%%%%%%%%%%%%%%%%%%%%%%

%%%%%%%%%%%%%%%%%%%%%%%%%%%%%%%%%%%%%%%%%%%%%%%%%%%%%%%%%%%%%%%%%%%%%%%%%%%%%%%

\end{document}